\documentclass[iop]{emulateapj}
\slugcomment{{\sc Submitted to ApJ Supplement: 3 November 2017}}
\usepackage{amssymb,amsmath,bm}
\usepackage{graphicx}
\usepackage{natbib}
\usepackage{adjustbox}
\usepackage{dblfnote}
\usepackage[colorlinks=true, linkcolor=blue,urlcolor=blue,citecolor=blue]{hyperref}

\begin{document}

\title{Merging Cluster Collaboration: Optical and Spectroscopic Survey of a Radio-Selected Sample of Twenty Nine Merging Galaxy Clusters} 
\shorttitle{Galaxy Survey of Radio Relic Clusters}

\author{N. Golovich*\altaffilmark{1,2},
W. A. Dawson\altaffilmark{2},
D. M. Wittman\altaffilmark{1,3}\\
M. J. Jee\altaffilmark{1,4},
B. Benson\altaffilmark{1},
B. Lemaux\altaffilmark{1},
R. J. van Weeren\altaffilmark{5},
F. Andrade-Santos\altaffilmark{5},
D. Sobral\altaffilmark{6,7},
F. de Gasperin\altaffilmark{6,8},
M. Br\"{u}ggen\altaffilmark{8},
M. Brada\v{c}\altaffilmark{1},
K. Finner\altaffilmark{4},
A. Peter\altaffilmark{9,10,11}}


\altaffiltext{1}{Department of Physics, University of California, One Shields Avenue, Davis, CA 95616, USA}
\altaffiltext{2}{Lawrence Livermore National Laboratory, 7000 East Avenue, Livermore, CA 94550, USA}
\altaffiltext{3}{Instituto de Astrof\'{\i}sica e Ci\^{e}ncias do Espa\c{c}o, Universidade de Lisboa, Lisbon, Portugal}
\altaffiltext{4}{Department of Astronomy, Yonsei University, 50 Yonsei-ro, Seodaemun-gu, Seoul, South Korea}
\altaffiltext{5}{Harvard-Smithsonian Center for Astrophysics, 60 Garden Street, Cambridge, MA 02138, USA; Clay Fellow}
\altaffiltext{6}{Leiden Observatory, Leiden University, P.O. Box 9513, 2300 RA Leiden, the Netherlands}
\altaffiltext{7}{Department of Physics, Lancaster University, Lancaster, LA1 4YB, UK}
\altaffiltext{8}{Hamburger Sternwarte, Universit\"at Hamburg, Gojenbergsweg 112, 21029 Hamburg, Germany}
\altaffiltext{9}{Department of Astronomy, The Ohio State University, 140 W. 18th Avenue, Columbus, OH 43210, USA}
\altaffiltext{10}{Center for Cosmology and AstroParticle Physics, The Ohio State University, 191 W. Woodruff Avenue, Columbus, OH 43210, USA}
\altaffiltext{11}{Department of Physics, The Ohio State University, 191 W. Woodruff Avenue, Columbus, Ohio 43210, USA}

\email{* nrgolovich@ucdavis.edu\\}
\shortauthors{Golovich et al.}

\label{firstpage}
\begin{abstract}
Multi-band photometric and multi-object spectroscopic surveys of merging galaxy clusters allow for the characterization of the distributions of constituent dark matter and galaxy populations, constraints on the dynamics of the merging subclusters, and an understanding of galaxy evolution of member galaxies. We present deep photometric observations from Subaru/SuprimeCam and a catalog of $\sim$5400 spectroscopic cluster members from Keck/DEIMOS across 29 merging galaxy clusters ranging in redshift from $z=0.07$ to $0.55$. The ensemble is compiled based on the presence of radio relics, which highlight cluster scale collisionless shocks in the intra-cluster medium. Together with the spectroscopic and photometric information, the velocities, timescales, and geometries of the respective merging events may be tightly constrained. In this preliminary analysis, the velocity distributions of 28 of the 29 clusters are shown to be well fit by single Gaussians. This indicates that radio relic mergers largely occur transverse to the line of sight and/or near apocenter. In this paper, we present our optical and spectroscopic surveys, preliminary results, and a discussion of the value of radio relic mergers for developing accurate dynamical models of each system. 
\end{abstract}

\begin{keywords} 
{Galaxies: clusters: general---dark matter---galaxies:  evolution---shock waves}
\end{keywords}


\section{Introduction}\label{sec:intro}

Merging galaxy clusters have been established as fruitful astrophysical laboratories. In particular, `dissociative mergers' \citep{Dawson:2012}, where two galaxy clusters have collided and the effectively collisionless galaxies and dark matter have become dissociated from the collisional ICM which collides and slows during the merger, are a particularly interesting subclass of mergers. They have been used to place tight constraints on the dark matter self-interaction cross-section \citep[DM; e.g.,][]{Clowe06,Randall:2008}, understand fundamental particle/plasma physics associated with the intra-cluster medium \citep[ICM; e.g.,][]{Blandford:1987,Markevitch:2001b,Brunetti:2014,vanWeeren:2017}, and merger related galaxy evolution \citep[e.g.,][]{Miller:2003,Poggianti:2004,Chung:2009,Stroe:2014a,Stroe:2017,Mansheim:2017a,Mansheim:2017b}. These studies have allowed for new and broader understanding of the content, distribution, and interactions between and within each component. These studies are complicated by: the complexity of the merger properties (mass, dynamics, etc.), the range of disparate observations necessary to form a synoptic understanding of any one merger, and the limited sample size of dissociative mergers to study.

Mergers are complex physical phenomena, where dynamical parameters such as the merger speed at pericenter, the elapsed time since pericenter, and the merger geometry are typically unknown. This leaves a vast volume of parameter space that must be considered in any subsequent analysis to properly propagate uncertainty \citep[e.g.,][]{Lage:2015}. The volume of parameter space that must be explored can be shrunk by studying the separate components of the merger as a whole \citep[e.g.,][]{Dawson:2013,Ng:2015,Golovich:2016}.

Observationally, each component of a merger is probed differently. The DM must be inferred using gravitational lensing techniques that necessitate deep photometric images \citep[see][for a review]{Bartelmann:2001, Hoekstra:2013}. The ICM is hot ($\sim$ several keV) and emits thermal bremsstrahlung X-rays \citep[e.g.,][]{Cavaliere:1976} which may be observed spatially and spectroscopically with modern X-ray observatories in orbit. Non-thermal emission from the ICM may be observed with radio telescopes, which reveals complex microphysics of particle acceleration and turbulence \citep[see e.g.,][]{Brunetti:2008}. The galaxies may be observed photometrically and spectroscopically. Photometry in multiple bands allow for semi-precise photometric redshift estimates \citep[see e.g.,][]{Benitez:2000,Bolzonella:2000} and red sequence selection of cluster members \citep[e.g.,][]{Kodama:1997}. Spectroscopic observations, on the other hand, allow for precise redshift estimation, but these observations are much more expensive and usually result in incomplete surveys of member galaxies. Spectroscopy may also be used to study the effects of the cluster environment on the constituent galaxies via line ratios \citep[e.g.,][]{BPT} including AGN and star formation rate studies \citep[e.g.,][]{Moore:1996,Miller:2003,Stroe:2014a,Sobral:2015}.

Circa 2012 all dissociative mergers were identified and confirmed using an array of the aforementioned observations. Collecting and analyzing this array of observations was resource intensive, which in large part is the reason for the small sample of dissociative mergers \citep[see][for a list of the eight known dissociative mergers circa 2012]{Dawson:2012}.
In recent years, we have implemented new technique of quickly identifying dissociative merging galaxy clusters via detection of enhanced, diffuse radio emission has become fruitful. Radio relics and radio halos appear in radio images between $\sim$100 MHz and several GHz as Mpc-scale, diffuse radio features. They are thought to trace synchrotron emission from electrons interacting with shocks and turbulent motion \citep[e.g.,][]{Brunetti:2008,Feretti:2012}, and thus should be associated with dissociative mergers. Magneto-hydrodynamical simulations of cluster mergers confirm this, and can reporduce key features of radio relics \citep[e.g.,][]{Skillman:2013, Vazza:2016}.
Because radio relic selection of dissociative mergers can be done with a single band wide field survey, while maintaining a high purity (as demonstrated in this paper), it is more economical compared to previous multi-probe selection methods. 

Spectroscopic and photometric observations of the galaxies of merging subclusters allow for estimation of the dynamical properties of individual merging systems. We have demonstrated this with a series of studies of individual merger systems \citep[CIZA J2242.8+5301, El Gordo, MACS J1149.5+2223, ZwCl 0008.8+5215, Abell 3411, and ZwCl 2341.1+0000 presented in][respectively]{Dawson:2015,Ng:2015,Golovich:2016,Golovich:2017,vanWeeren:2017,Benson:2017}. The dynamical models of individual clusters greatly reduce the vast parameter space that simulators must explore to reproduce underlying astrophysics. The presence of radio relics in each of these systems has been shown to greatly improve the precision of dynamical models \citep{Ng:2015,Golovich:2016,Golovich:2017}, and direct study of the underlying shock and radio relics have yielded insight into particle acceleration models \citep[e.g.,][]{Brunetti:2014, vanWeeren:2017}.

In this paper, we outline our photometric and spectroscopic observations of an ensemble of 29 radio relic mergers. In \S\ref{sec:sample} we describe the construction of the ensemble of 29 merging systems. In \S\ref{sec:overview} we detail our photometric and spectroscopic observational campaign including the technical details of the observations, data reduction, and data processing. We compile and analyze the redshift global redshift distributions of each system in \S\ref{sec:analysis}, and we discuss the implications of radio selection and offer conclusions in \S\ref{sec:discussion}. 

We assume a flat $\Lambda$CDM universe with $H_0 = 70\,\mathrm{km}\,\mathrm{s}^{ -1}\,\mathrm{Mpc}^{-1}$, $\Omega_M = 0.3$, and $\Omega_\Lambda = 0.7$. AB magnitudes are utilized throughout, and all distances are proper. 


\section{Radio Relic Sample}\label{sec:sample}

Constraints on the dark matter self-interaction cross-section is one of the driving science cases for this survey.
A radio relic selection has a number of potential advantages for this science case over other selection methods:
(1) it guarantees against the selection of pre-pericentric systems since the presence of a radio relic indicates a shockwave traveling in the ICM due to a major merger;
(2) it will disfavor the very youngest post-pericentric systems, which have not had time to generate radio relics, and where the offset between the effectively collisionless galaxies and potentially self-interacting dark matter has not had a chance to increase to a potential maximal offset \citep{Kahlhoefer:2013};
(3) it is biased towards selecting mergers in the plane of the sky where any observable offset between the galaxies, dark matter, and ICM will be maximized (this is also important for other astrophysical studies) \citep{ensslin1998}; and,
(4) as noted in \S1, a large sample of dissociative mergers can be prudently compiled.

The first detection of a radio relic in a merging galaxy cluster was in the Coma Cluster \citep{Ballarati:1981}. Radio relics were subsequently discovered individually through pointed observations of known merging systems. In the last decade, searches of wide-area radio surveys have increased the rate of detection \citep[e.g.,][]{vanWeeren:2011}. Several potential radio relics were discovered through comparisons of radio surveys with the ROSAT All-Sky Survey catalogs \citep[RASS:][]{RASS}. Follow up of these objects resulted in several discoveries \citep{vanWeeren:2009,vanWeeren:2009b,vanWeeren:2010,vanWeeren:2011c,vanWeeren:2012,vanWeeren:2012b,vanWeeren:2013}. Our sample begins with these radio relics, along with additional radio relics known by September 2011 listed in Table 3 of \citet{Feretti:2012}. Each of these clusters contain low surface brightness, steep-spectrum, polarized, and extended radio sources that lie at the periphery of the cluster (for individual observational papers see references therein). Relics classified as having a round morphology were discarded since they are likely radio phoenixes rather than Mpc scale cluster shocks. Radio phoenixes are generally associated with aged radio galaxy lobes that are re-energized through compression or other mechanisms \citep[e.g.,][]{deGasperin:2015b}. We imposed cuts designed to enable spectroscopic and weak lensing followup. Systems at very low redshift are not efficient lenses, so we eliminate clusters at $z<0.07$. We also eliminate systems not observable from the Mauna Kea observatories ($\delta<-31\degr$) from which we were awarded observational time. To this list, we added three additional radio relic systems that have appeared in the literature and pass the same selection criteria \citep[MACS J1149.5+2223, PSZ1 G108.18-11.5 and ZwCl 1856+6616, hereafter MACSJ1149, PSZ1G108 and ZwCl1856, respectively:][]{Bonafede:2012,deGasperin:2014,deGasperin:2015}. Finally, we added one of the radio phoenix relics (Abell 2443, hereafter A2443) to our spectroscopic survey due to a gap in an observing run. In total, our sample contains 29 systems listed in Table~\ref{tab:sample}.

\begin{table*}
\begin{center}
\caption{The Merging Cluster Collaboration radio-selected sample.}
\begin{tabular}{llllll}
Cluster				&	Short name 	& RA 		&	DEC 		&	Redshift	&	Discovery band		 \\
\hline 
1RXS J0603.3+4212		& 	1RXSJ0603	& 06:03:13.4 	&	+42:12:31 	&	0.226	&	Radio\\
Abell 115				& 	A115			& 00:55:59.5 	&	+26:19:14 	&	0.193	&	Optical	\\
Abell 521 				&	A521			& 04:54:08.6 	&	-10:14:39 		&	0.247	&	Optical	\\
Abell 523 				&	A523			& 04:59:01.0	&	+08:46:30 	&	0.104	&	Optical	\\
Abell 746 				&	A746			& 09:09:37.0	&	+51:32:48 	&	0.214	&	Optical	\\
Abell 781 				&	A781			& 09:20:23.2	&	+30:26:15 	&	0.297	&	Optical	\\
Abell 1240		 	&	A1240		& 11:23:31.9	&	+43:06:29 	&	0.195	&	Optical	\\
Abell 1300 			&	A1300		& 11:32:00.7	&	-19:53:34 		&	0.306	&	Optical \\
Abell 1612 			&	A1612		& 12:47:43.2	&	-02:47:32 		&	0.182	&	Optical	\\
Abell 2034 			&	A2034		& 15:10:10.8	&	+33:30:22		&	0.114	&	Optical	\\
Abell 2061 			&	A2061		& 15:21:20.6	&	+30:40:15 	&	0.078	&	Optical	\\
Abell 2163 			&	A2163		& 16:15:34.1	&	-06:07:26 		&	0.201	&	Optical	\\
Abell 2255 			&	A2255		& 17:12:50.0	&	+64:03:11 	&	0.080	&	Optical	\\
Abell 2345 			&	A2345		& 21:27:09.8	&	-12:09:59 		&	0.179	&	Optical	\\
Abell 2443 			&	A2443		& 22:26:02.6	&	+17:22:41 	&	0.110	&	Optical	\\
Abell 2744 			&	A2744		& 00:14:18.9	& 	-30:23:22 		&	0.306	&	Optical \\
Abell 3365 			&	A3365		& 05:48:12.0	& 	-21:56:06 		&	0.093	&	Optical\\
Abell 3411 			&	A3411		& 08:41:54.7	& 	-17:29:05 		&	0.163	&	Optical\\
CIZA J2242.8+5301		& 	CIZAJ2242	& 22:42:51.0	& 	+53:01:24 	&	0.189	&	X-ray\\
MACS J1149.5+2223	& 	MACSJ1149	& 11:49:35.8	& 	+22:23:55 	&	0.544	&	X-ray\\
MACS J1752.0+4440	&	MACSJ1752	& 17:52:01.6	& 	+44:40:46 	&	0.365	&	X-ray\\
PLCKESZ G287.0+32.9 	&	PLCKG287	& 11:50:49.2	& 	-28:04:37 		&	0.383	&	SZ \\
PSZ1 G108.18-11.53 	&	PSZ1G108	& 23:22:29.7	&	 +48:46:30	&	0.335	&	SZ \\
RXC J1053.7+5452 		&	RXCJ1053	& 10:53:44.4	& 	+54:52:21 	&	0.072	&	X-ray \\
RXC J1314.4-2515 		&	RXCJ1314	& 13:14:23.7	& 	-25:15:21 		&	0.247	&	X-ray \\
ZwCl 0008.8+5215 		&	ZwCl0008		& 00:08:25.6	&	+52:31:41 	&	0.104	&	Optical\\
ZwCl 1447+2619 		&	ZwCl1447		& 14:49:28.2	&	+26:07:57 	&	0.376	&	Optical	\\
ZwCl 1856.8+6616		& 	ZwCl1856		& 18:56:41.3	&	+66:21:56.0	&	0.304	&	Optical	\\
ZwCl 2341+0000 		&	ZwCl2341		& 23:43:39.7	&	+00:16:39 	&	0.270	&	Optical	\\
\end{tabular}
\label{tab:sample}
\end{center}
\end{table*}

The sample is predominantly composed of low redshift ($\sim$0.1--0.3) clusters due to radio relics typically being discovered in wide, shallow surveys \citep[e.g. NVSS:][]{Condon:1998}. This is a reasonable redshift range for lensing follow-up, and also has the advantage of mapping given physical separations to substantial angular separations for spatial analysis of cluster components. 

\begin{figure*}
\begin{center}
\includegraphics[width=\textwidth]{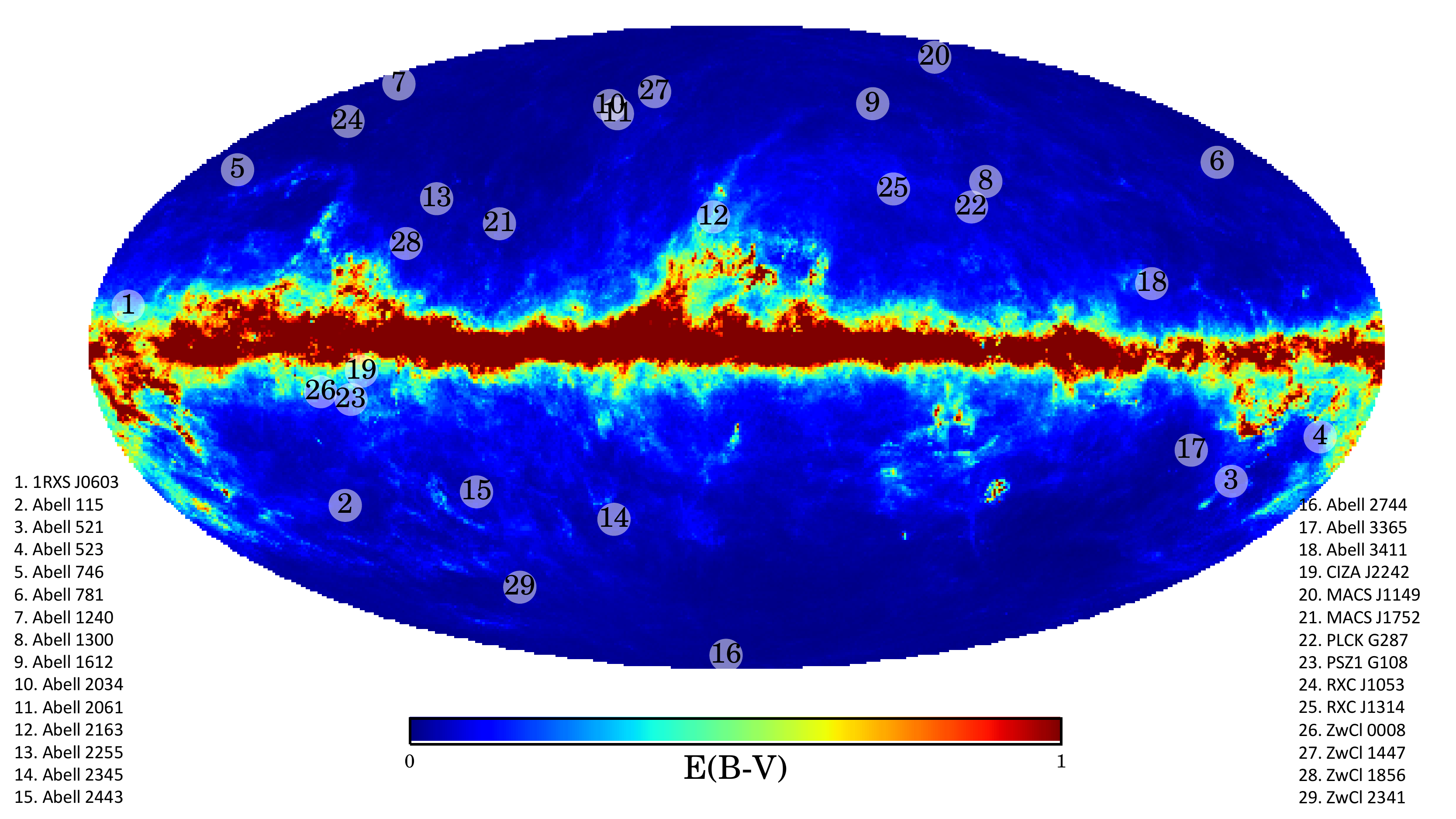}
\caption{Galactic dust extinction map \citep{Schlegel:1998} with overlaid positions of the 29 systems in our sample.}
\label{fig:dust}
\end{center}
\end{figure*}

The radio selection strategy brings challenges in terms of obtaining spectroscopy and lensing followup. Because radio surveys have gone right through the galactic plane, many of the systems suffer more extinction than is typical in visible-wavelength surveys. The all sky galactic dust extinction map is presented in Figure \ref{fig:dust} with all 29 systems in our sample. The most extreme example is CIZAJ2242 with $A_V \approx 1.4$ \citep[the approximation sign emphasizes that the extinction varies over the field;][]{Schlegel:1998}. \citet{Dawson:2015} describes the success of the position-dependent extinction corrections applied to that system in terms of yielding uniform color selection of cluster members, and \citet{Jee:2015} demonstrates that weak lensing can be efficiently measured despite the extinction. The low galactic latitude also affected the spectroscopy not only through extinction but also by causing more slits to be wasted on stars. A contributing factor in some cases was the poor quality of imaging available at the time of slit mask design. Blended binary stars were not rejected in morphological cuts, and constituted a substantial contamination. The next most extincted systems in the sample are A2163 and ZwCl0008, for which $A_V\sim 0.8$. We therefore expect the lensing and galaxy analyses of most of the systems in this sample to exceed the quality of those for CIZAJ2242. We have corrected our photometry for extinction throughout. 

The resulting 29 systems are listed in Table \ref{tab:sample}. For each system, the following milestones are to be achieved for each cluster: 

\begin{itemize}
\item Observations including spectroscopic, ground based wide field photometric, space based pointed photometric, X-ray and radio
\item Optical analysis to estimate the number and location of subclusters 
\item Redshift analysis to estimate line of sight velocity information of subclusters
\item X-ray and radio analysis of shocks and radio relics including polarization measurements
\item Weak lensing analysis to find location and mass of subclusters
\item Dynamical analysis 
\end{itemize}

In this paper, we will discuss the spectroscopic and wide-field optical observations for our sample of 29 merging clusters. These will ultimately result in two of the three primary inputs for the dynamics analysis as well as classify the mergers by their complexity and reasonability to probe astrophysical hypotheses including merging induced galaxy evolution, particle acceleration at cluster shocks, merger induced turbulence, and self-interacting dark matter models. The remaining goals will be achieved in follow up papers utilizing the data presented here. 

\section{Optical Imaging and Spectroscopic Observational Campaign}\label{sec:overview}

\subsection{Survey Goals and Requirements}
The goal of the optical imaging survey is to obtain lensing quality, wide-field imaging in at least two photometric bands. The two filters are chosen to straddle the 4000\AA$\,$break in order to select cluster members photometrically via red sequence relations. Furthermore, our weak lensing method makes use of these red sequence relations in order to select background galaxies for lensing studies \citep{Jee:2015,Jee:2016, Golovich:2017}. Additionally, for clusters that our SuprimeCam observations came before our DEIMOS observations, we made use of the SuprimeCam images for spectroscopic target selection (see \S\ref{sssec:KeckTargetSelection}). Many clusters have archival imaging that we have obtained. We observed 18 systems with Subaru/SuprimeCam to complete the photometric survey. 

The spectroscopic survey has a goal of obtaining $\sim 200$ member galaxy velocities in each system. We used redshifts from the literature when available in order to reduce the amount of new observations required. When obtaining new spectra, we designed observations to also meet the goal of enabling studies of recent star formation and ultimately the link between mergers and star formation. We achieve this by adjusting the observed wavelength range for each cluster to the emitted wavelength range from H$\beta$ to H$\alpha$ for clusters with $z \lesssim 0.3$ and [\ion{O}{2}] to [\ion{O}{3}] for clusters with $z \gtrsim 0.3$. The data available for star formation studies therefore varies from cluster to cluster depending on the number of previously published redshifts and the redshift of the cluster. Additional observations were required for 18 systems with many having no more than a handful of previously published member redshifts. In the following subsections we will detail the targeting, observing, and data reductions of our optical and spectroscopic surveys.

\subsection{Subaru/SuprimeCam Observations}\label{subsec:subaru_observations}

We observed 18 clusters over four nights using the 80 Megapixel SuprimeCam \citep{SuprimeCam} camera on the Subaru Telescope on Mauna Kea. Table \ref{tab:optical_survey} summarizes these observations. The basic strategy is to achieve weak lensing quality in one filter and obtain a second filter to define the color of detected objects by straddling the 4000 \AA$\,$break. For the lensing quality image, the exposure time was 2880 s (8$\times$360 s) and we rotated the field between each exposure by 15$\degr$ in order to distribute the bleeding trails and diffraction spikes from bright stars azimuthally to be later removed by median-stacking. This scheme enabled us to maximize the number of detected galaxies, especially for background source galaxies for weak lensing near stellar halos or diffraction spikes. In the second and third filters (g and/or i), the exposure time was 720 s (4$\times$180 s). These exposures were rotated by 30$\degr$ from exposure to exposure for the same reason as above. In order to efficiently fill the time of each observing night, we added a third band to several clusters. The actual observing times may vary due to real-time changes to the observational plan due to unexpected lost time. 

\begin{table*}[h!]
\begin{center}
\small
\caption{Merging Cluster Collaboration radio-relic selected Subaru/SuprimeCam survey.}
\begin{tabular}{lllll}
Cluster 					&	Filter		&	Date					&	Seeing (arcsec)		&	Exposure (s)\\
\hline
1RXS J060313.4+421231		&	g		&	2014 February 25		&	0.57				&	720	\\
1RXS J060313.4+421231		&	r		&	2014 February 25		&	0.57				&	2880\\
1RXS J060313.4+421231		&	i		&	2014 February 25		&	0.50				&	720	\\
Abell 523					&	g		&	2014 February 26		&	1.00				&	720	\\
Abell 523					&	r		&	2014 February 26		&	0.78				&	2880\\
Abell 746					&	g		&	2014 February 26		&	0.88				&	720	\\
Abell 746					&	r		&	2014 February 26		&	1.01				&	2880\\
Abell 1240				&	g		&	2014 February 25		&	0.67				&	720	\\
Abell 1240				&	r		&	2014 February 25		&	0.58				&	2880\\
Abell 1300				&	g		&	2014 February 26		&	0.89				&	720	\\
Abell 1300				&	r		&	2014 February 26		&	0.88				&	2160	\\
Abell 2061				&	g		&	2013 July 13			&	0.68				&	720\\
Abell 2061				&	r		&	2013 July 13			&	0.67				&	2520	\\
Abell 2061				&	i		&	2013 July 13			&	0.60				&	2676	\\
Abell 2061				&	i		&	2014 February 26		&	0.65				&	720  \\
Abell 3365				&	g		&	2014 February 25		&	0.97				&	720 \\
Abell 3365				&	r		&	2014 February 25		&	0.71				&	2880	\\
Abell 3365				&	i		&	2014 February 25		&	0.62				&	720 \\
Abell 3411					&	g		&	2014 February 25		&	0.80				&	720	\\
Abell 3411					&	r		&	2014 February 25		&	0.82				&	2880\\
Abell 3411					&	i		&	2014 February 25		&	0.77				&	720	\\
CIZA J2242.8+5301			&	g		&	2013 July 13			&	0.63				&	720	\\
CIZA J2242.8+5301			&	i		&	2013 July 13			&	0.55				&	3400	\\
MACS J175201.5+444046	&	g		&	2013 July 13			&	0.62				&	720	\\
MACS J175201.5+444046	&	r		&	2013 July 13			&	0.64				&	1440\\
MACS J175201.5+444046	&	i		&	2013 July 13			&	0.63				&	2520	\\
MACS J175201.5+444046	&	i		&	2014 February 26		&	0.73				&	1260	\\
PLCKESZ G287.0+32.9		&	g		&	2014 February 26		&	0.81				&	720	\\
PLCKESZ G287.0+32.9		&	r		&	2014 February 26		&	0.97				&	2880\\
PSZ1 G108.18-11.53		&	g		&	2015 September 12		&	0.65				&	1440 \\
PSZ1 G108.18-11.53		&	r		&	2015 September 12		&	0.55				&	2520\\
RXC J1053.7+5452			&	g		&	2014 February 26		&	0.83				&	720	\\
RXC J1053.7+5452			&	r		&	2014 February 26		&	0.92				&	720	\\
RXC J1314.4-2515			&	g		&	2014 February 25		&	0.86				&	720\\
RXC J1314.4-2515			&	r		&	2014 February 25		&	0.71				&	2880	\\
RXC J1314.4-2515			&	NB814	&	2014 February 25		&	0.77				&	1000	\\
ZwCl 0008.8+5215			&	g		&	2013 July 13			&	0.52				&	720\\
ZwCl 0008.8+5215			&	r		&	2013 July 13			&	0.57				&	2880	\\
ZwCl 1447+2619			&	g		&	2014 February 26		&	0.91				&	720	\\
ZwCl 1447+2619			&	r		&	2014 February 26		&	0.76				&	2880\\
ZwCl 1447+2619			&	i		&	2014 February 26		&	0.55				&	720	\\
ZwCl 1856+6616			&	g		&	2015 September 12		&	0.70				&	720 \\
ZwCl 1856+6616			&	r		&	2015 September 12		&	0.65				&	2520\\
ZwCl 2341+0000			&	g		&	2013 July 13			&	0.49				&	720	\\
ZwCl 2341+0000			&	r		&	2013 July 13			&	0.50				&	2880\\
\end{tabular}
\label{tab:optical_survey}
\end{center}
\end{table*}

Archival Subaru/SuprimeCam imaging was downloaded from the SMOKA data archive \citep{SMOKA}, and is detailed in Table \ref{tab:SC_archive}. We note that the observational strategy for the archival data did not prescribe rotating between exposures, so diffraction spikes and bleeding trails are present. Also, we did not make use of the full set of archival images for these clusters since we only required two bands of imaging in order to define the color and complete a color--magnitude selection. We utilized the deepest images available that satisfy this requirement ensuring good seeing conditions. 

\begin{table*}
\begin{center}
\caption{Archival imaging from Subaru/SuprimeCam utilized in this study.}
\begin{tabular}{lllll}
Cluster 					&	Filter		&	Date									&	Exposure (s)\\
\hline
Abell 115					&	V		&	2003 September 25, 2005 October 03		&	1530	\\
Abell 115					&	i		&	2005 October 03						&	2100\\
Abell 521					&	V		&	2001 October 14						&	1800\\
Abell 521					&	R		&	2001 October 15						&	1620\\
Abell 781					&	V		&	2010 March 14, 15						&	3360\\
Abell 781					&	i		&	2010 March 15							&	2160\\
Abell 2034				&	g		&	2005 April 11							&	720\\
Abell 2034				&	R		&	2005 April 11, 2007 June 19				&	12880\\
Abell 2163				&	V		&	2009 April 30							&	2100\\
Abell 2163				&	R		&	2008 April 07							&	4500\\
Abell 2255				&	B		&	2007 August 14							&	1260\\
Abell 2255				&	R		&	2007 August 14							&	2520\\
Abell 2345				&	V		&	2010 June 10, 2010 November 10			&	3600\\
Abell 2345				&	i		&	2005 October 03						&	2100\\
Abell 2744				&	B		&	2013 July 16							&	2100\\
Abell 2744				&	R		&	2013 July 15							&	3120\\
MACS J1149				&	V		&	2003 April 05							&	2520\\
MACS J1149				&	R		&	2003 April 05, 2005 March 05, 2010 March 18	&	5490\\
\end{tabular}
\label{tab:SC_archive}
\end{center}
\end{table*}

\subsubsection{Subaru/SuprimeCam: Data Reduction}\label{sssec:subaru_reduction}

\begin{figure*}[ht!]
\begin{center}
\includegraphics[width=\textwidth]{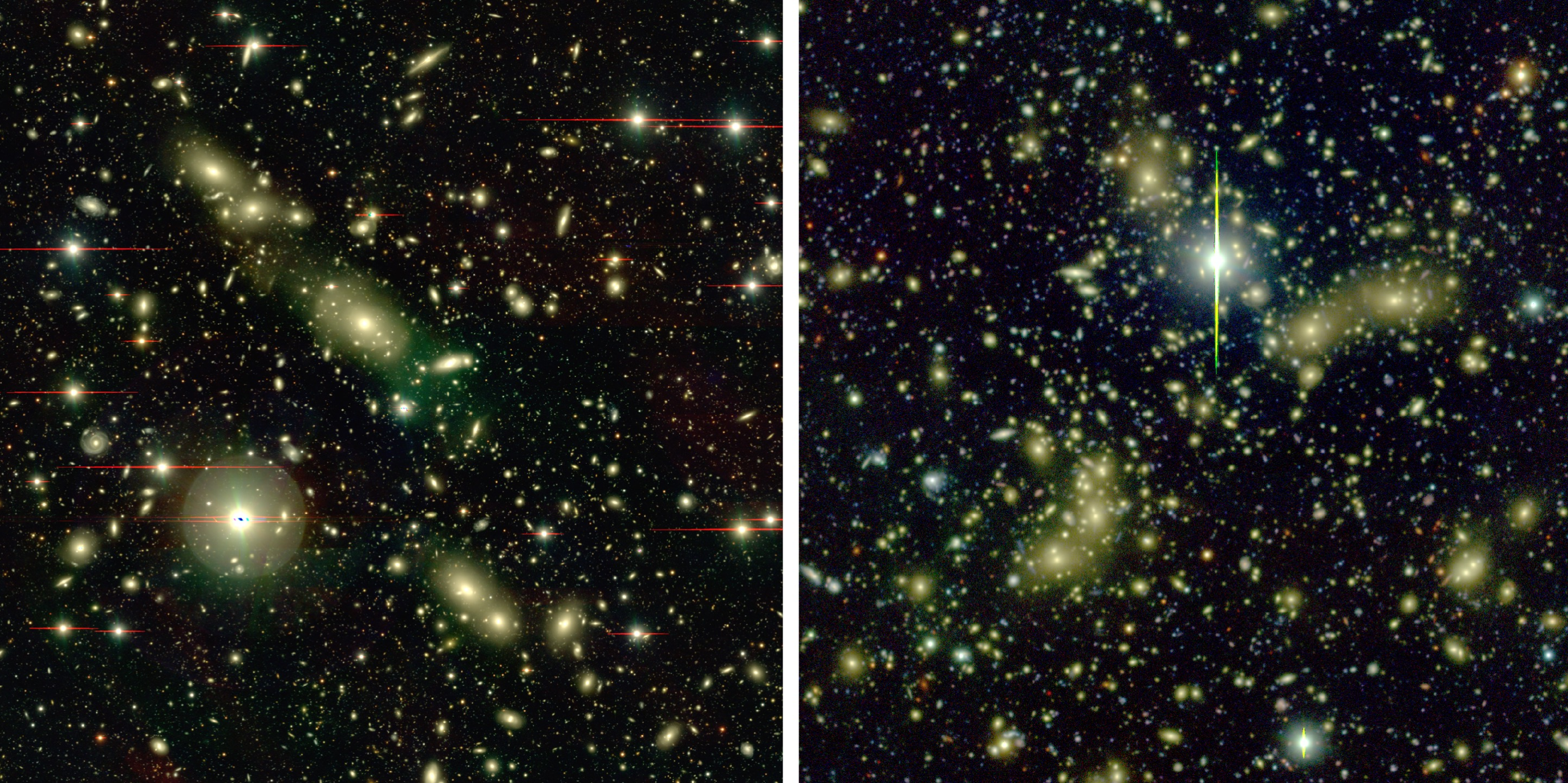}
\caption{Example Subaru/SuprimeCam images of the central regions of A2061 (left) and A2744 (right). A2061 is displayed using our g, r, and i band images while A2744 is displayed using archival B, R, and Z images. Note that the i-band image for A2061 and the Z-band image for A2744 were used only to make these true-color images. These images were combined using the trilogy software \citep{Coe:2012}.}
\label{fig:subaru}
\end{center}
\end{figure*}

The CCD processing (overscan subtraction, flat-fielding, bias correction, initial geometric distortion rectification, etc) were carried out with the SDFRED2 package \citep{SDFRED2}. Much of the archival data required the first version of this pipeline \citep[SDFRED1:][]{SDFRED1}. We refine the geometric distortion and World Coordinate System (WCS) information using the SCAMP software \citep{SCAMP}. The Two Micron All Sky Survey \citep[2MASS; ]{2mass} catalog was selected as a reference when the SCAMP software was run except for clusters covered by the Sloan Digital Sky Survey \citep[SDSS][]{sdss5}, for which the Data Release 5 catalogs were used. We also rely on SCAMP to calibrate out the sensitivity variations across different frames. For image stacking, we ran the SWARP software \citep{SWARP} using the SCAMP result as input. We first created median mosaic images and then used it to mask out pixels (3$\sigma$ outliers) in individual frames. These masked frames were weight-averaged to generate the final mosaic, which is used for the scientific analyses hereafter. Two example images are presented in Figure \ref{fig:subaru}.

\subsubsection{Subaru/SuprimeCam: Photometric Catalog Generation}\label{ssec:photometry}

Object detection is achieved with Source Extractor \citep{sextractor} in dual image mode using the deepest image for detection. The blending threshold parameter {\tt BLEND-NTHRESH} is set to 32 with a minimal contact {\tt DEBLEND}$\_${\tt MINCONT} of 10$^{-\text{4}}$. We employ reddening values from \cite{Schlafly:2011} to correct for dust extinction, which are listed in Tables \ref{tab:optical_survey} and \ref{tab:SC_archive}. Zero points were transferred from SDSS for the overlapping clusters and transferred to the clusters outside the SDSS footprint observed on the same night with SuprimeCam accounting for atmospheric extinction related to the airmass differences of our observations. Atmospheric extinction values for Mauna Kea were taken from \cite{Buton:2013}. 

Since the sample has relatively low redshift, it is expected for cluster members to have high S/N and correspondingly good photometry. We enforce that potential cluster member objects have uncertainties in their magnitudes of less than 0.5 magnitudes, and we remove all objects brighter than the BCG, which we have identified spectroscopically in each cluster. These cuts eliminate most bright foreground galaxies and stars as well as false detections at extremely faint magnitudes. Only objects within $R_{200}$ \citep[as determined from our redshift analysis and scaling relations][]{Evrard:2008,Duffy} of the center of the cluster are retained. This limits the vignetting of the edges as well as removes spurious detections near the edge of the field.

\subsection{Keck/DEIMOS Observations}\label{subsec:DEIMOS_observations}

\begin{figure*}[t!]
\begin{center}
\includegraphics[width=\textwidth]{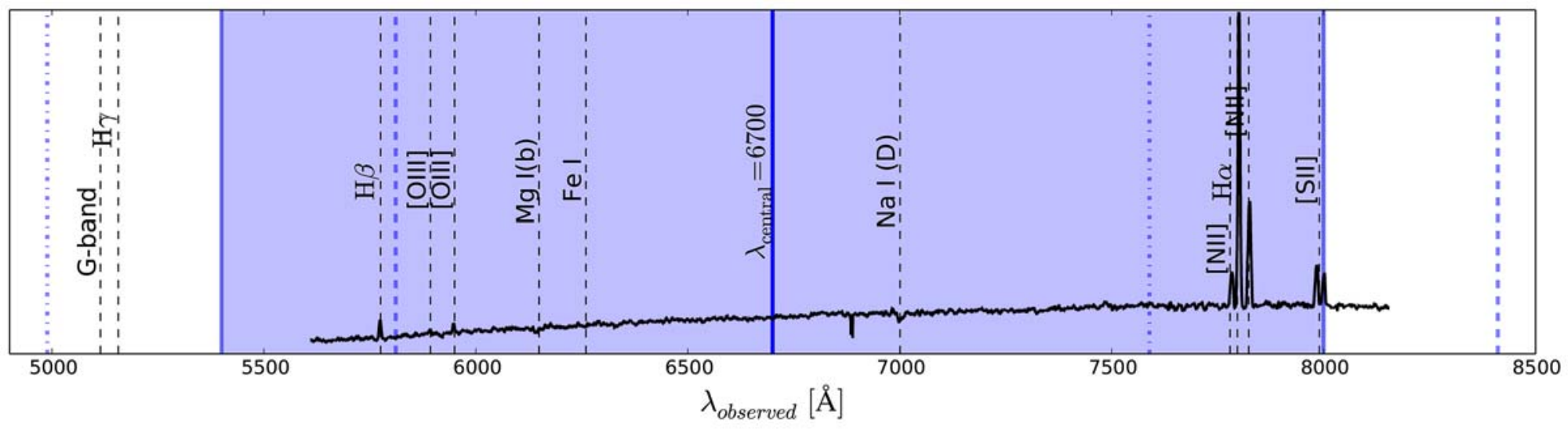}
\caption{{\it Reprinted} Figure 2 of \citet{Dawson:2015}. Example spectral coverage of the Keck/DEIMOS observations (shaded blue region) for a low redshift ($z\leq0.3$) cluster, along with the redshifted location of common cluster emission and absorption features (black dashed lines). The blue dot-dash pair and the blue dashed pair of lines show the variable range depending on where the slit was located along the width of the slit mask. The solid black line shows an example galaxy spectrum from our DEIMOS survey.}
\label{fig:SpecCoverage}
\end{center}
\end{figure*}

We conducted a spectroscopic survey utilizing the DEIMOS multi-object spectrograph \citep{DEIMOS} on the Keck II telescope at the W. M. Keck Observatory on Mauna Kea over the following nights: 26 January 2013, 14 July 2014, 5 September 2014, 3-5 December 2013 (half nights), 22-23 June 2014, 15 February 2015, and 13 December 2015. In total, 54 slit masks were observed. Each was milled with 1$\arcsec$\ wide slits and utilized the 1200\,line\,mm$^{-1}$ grating, which results in a pixel scale of $0.33$\,\AA\,pixel$^{-1}$ and a resolution of $\sim1$\,\AA\ (50\,km\,s$^{-1}$). For clusters with a redshift below 0.3, the grating was tilted to observe the following spectral features:  H$\beta$, [\ion{O}{3}], \ion{Mg}{1} (b), \ion{Fe}{1},  \ion{Na}{1} (D), [\ion{O}{1}], H$\alpha$, and the [\ion{N}{2}] doublet. A typical wavelength coverage of 5400\,\AA\ to 8000\,\AA\, is shown in Figure \ref{fig:SpecCoverage} for a galaxy observed in CIZAJ2242. The actual wavelength coverage may be shifted by $\sim\pm400$\AA\, depending where the slit is located along the width of the slit mask. This spectral setup enables us to also study the star formation properties of the cluster galaxies; see related work by \citet{Sobral:2015}. For higher redshift clusters (above 0.3), the grating was tilted to instead cover the following spectra features:   [\ion{O}{2}], Ca(H), Ca(K), H$\delta$, G-band, H$\gamma$, H$\beta$, and [\ion{O}{3}]. The position angle (PA) of each slit was chosen to lie between $\pm5\degr$ to 30$\degr$ of the slit mask PA to achieve optimal sky subtraction during reduction with the DEEP2 version of the spec2d package \citep{DEEP2:2013}. In general, for each mask we took three $\sim$900\,s exposures except for a few cases where a few extra minutes at the end of the night were spent on an individual mask or when weather altered our observation plans in the middle of the night. In total, 54 slit masks were observed with a total of $\sim$7000 slits over the course of the spectroscopic survey. 

\begin{table*}
\begin{center}
\scriptsize
\caption{Merging Cluster Collaboration radio-relic selected spectroscopic survey.}
\begin{tabular}{llllll}
Slitmask 		&	Date				&	Target Imaging	&	Exposure (s)	&	Wavelength (\AA)&	Slits\\
\hline
1RXSJ0603-1	&	2013 January 16	&	WFC		&	3000		&	6200		&	105\\
1RXSJ0603-2	&	2013 January 16	&	WFC		&	3000		&	6200		&	100\\
1RXSJ0603-3	&	2013 September 5	&	WFC		&	3600		&	7000		&	98\\
1RXSJ0603-4	&	2013 September 5	&	WFC		&	3600		&	7000		&	87\\
A115-1		&	2014 June 22		&	SDSS	&	2500		&	6900		&	176\\
A115-2		&	2014 June 23		&	SDSS	&	2400		&	6900		&	142\\
A523-1		&	2013 January 16	&	WFC		&	3000		&	6200		&	99\\
A523-2		&	2013 December 4	&	WFC		&	2700		&	6200		&	94\\
A523-3		&	2015 February 16	&	WFC		&	2700		&	6300		&	111\\
A746-1		&	2013 January 16	&	SDSS	&	3600		&	6200		&	110\\
A1240-1		&	2013 December 3	&	SDSS	&	2700		&	6850		&	120\\
A1240-2		&	2015 February 16	&	SDSS	&	2700		&	6820		&	164\\
A1612-1		&	2015 February 16	&	SDSS	&	1200		&	6750		&	186\\
A2034-1		&	2013 July 14		&	SDSS	&	2700		&	6700		&	158\\
A2443-1		&	2014 June 22		&	SDSS	&	2400		&	6400		&	153\\
A2443-2		&	2014 June 23		&	SDSS	&	2400		&	6400		&	163\\
A3365-1		&	2013 January 16	&	WFC		&	2700		&	6200		&	68\\
A3365-2		&	2013 January 16	&	WFC		&	2400		&	6200		&	66\\
A3365-3		&	2013 December 3	&	WFC		&	2700		&	6300		&	63\\
A3365-4		&	2015 February 16	&	SC		&	2700		&	6200		&	160\\
A3411-1		&	2013 December 3	&	WFC		&	2700		&	6650		&	132\\
A3411-2		&	2013 December 3	&	WFC		&	2700		&	6650		&	127\\
A3411-3		&	2013 December 4	&	WFC		&	2700		&	6650		&	128\\
A3411-4		&	2013 December 4	&	WFC		&	2700		&	6650		&	131\\
A3411-5		&	2015 December 13	&	SC		&	3600		&	6650		&	142\\
CIZAJ2242-1	&	2013 July 14		&	WFC		&	2700		&	6700		&	148\\
CIZAJ2242-2	&	2013 July 14		&	WFC		&	2700		&	6700		&	126\\
CIZAJ2242-3	&	2013 September 5	&	SC		&	2700		&	7000		&	90\\
CIZAJ2242-4	&	2013 September 5	&	SC		&	2700		&	7000		&	106\\
MACSJ1752-1	&	2013 July 14		&	SDSS	&	2700		&	6700		&	155\\
MACSJ1752-2	&	2013 July 14		&	SDSS	&	2700		&	6700		&	119\\
MACSJ1752-3	&	2013 September 5	&	SDSS	&	3600		&	7000		&	114\\
MACSJ1752-4	&	2013 September 5	&	SDSS	&	2700		&	7000		&	118\\
PLCKG287-1	&	2015 February 16	&	SC		&	3900		&	7950		&	207\\
PLCKG287-2	&	2015 February 16	&	SC		&	2700		&	7950		&	185\\
PLCKG287-3	&	2015 February 16	&	SC		&	2700		&	7950		&	193\\
PSZ1G108-1	&	2014 June 22		& 	DSS		&	1800		&	7400		&	198\\
PSZ1G108-2	&	2014 June 23		& 	DSS		&	1800		&	7650		&	168\\
RXCJ1053-1	&	2013 January 16	&	SDSS	&	2803		&	6200		&	113\\
RXCJ1053-2	&	2013 December 3	&	SDSS	&	2700		&	6200		&	84\\
RXCJ1053-3	&	2013 December 4	&	SDSS	&	2430		&	6200		&	98\\
RXCJ1314-1	&	2015 February 16	&	SC		&	2520		&	7120		&	196\\
RXCJ1314-2	&	2015 February 16	&	SC		&	2520		&	7120		&	207\\
ZwCl0008-1	&	2013 January 16	&	WFC		&	2063		&	6200		&	81\\
ZwCl0008-2	&	2013 July 14		&	WFC		&	2700		&	6700		&	81\\
ZwCl0008-3	&	2013 September 5	&	WFC		&	2700		&	7000		&	75\\
ZwCl0008-4	&	2013 September 5	&	WFC		&	3600		&	7000		&	73\\
ZwCl1447-1	&	2014 June 22		&	SDSS	&	1520		&	7850		&	149\\
ZwCl1447-2	&	2014 June 23		&	SDSS	&	1053		&	7850		&	138\\
ZwCl1856-1	&	2014 June 22		& 	DSS		&	1800		&	7400		&	150\\
ZwCl1856-2	&	2014 June 23		& 	DSS		&	1800		&	7400		&	101\\
ZwCl2341-1	&	2013 July 14		&	SDSS	&	2700		&	6700		&	130\\
ZwCl2341-2	&	2013 July 14		&	SDSS	&	2700		&	6700		&	131\\
ZwCl2341-3	&	2013 September 5	&	SDSS	&	2700		&	7000		&	148\\
\end{tabular}
\label{tab:spec}
\tablecomments{Target Imaging codes: WFC$=$Issac Newton Telescope Wide Field Camera presnted in \citet{Reinout_INT_maps}, SDSS$=$Sloan Digital Sky Survey \citep[e.g.,][]{sdss12}, DSS$=$Palomar Observatory Digitized Sky Survey \citep{DSS}, SC$=$Subaru/SuprimeCam imaging (see \S\ref{subsec:subaru_observations})}
\end{center}
\end{table*}

\subsubsection{Keck/DEIMOS: Target Selection}\label{sssec:KeckTargetSelection}
 
Our primary objective for the spectroscopic survey was to maximize the number of cluster member spectroscopic redshifts in order to detect merging substructure within $R_{200}$. For each slit mask, the best imaging data available were utilized. For one third of the clusters this was our own SuprimeCam imaging from our simultaneous wide field imaging survey (see \S\ref{subsec:subaru_observations}). In the cases where this was unavailable at the time of our spectroscopic survey planning, we used the next best imaging at our disposal. SDSS Data Release 5 catalogs were utilized \citep{sdss5} for ten of the clusters, and for six of the clusters, this was INT WFC data presented in \citep{Reinout_INT_maps}. For the remaining two clusters, Digitized Sky Survey \citep[DSS:][]{DSS} imaging was utilized. For all imaging except the SDSS data, for which a photometric redshift selection was employed, a red sequence technique was utilized to select likely cluster members to create a galaxy number density map. The slit masks were then oriented to maximize the number of cluster members in the high red sequence density regions. Priors from the literature were also utilized in the placement of slit masks (e.g. lensing maps, X-ray surface brightness, radio relics, etc). 

The DEIMOS $5\arcmin\times 16.7\arcmin$ field-of-view (FOV) is very well suited to survey the low-z, elongated merging systems in our sample. In most cases, we aligned the long axis of our slit masks with the long axis of the system. The success of star-galaxy separation in our targeting data was variable and depended on the seeing of the imaging; thus, several of our slit masks were highly contaminated with stars. For example, for CIZAJ2242, which sits near the plane of the galaxy, has a stellar density nearly three times that of cluster members. When selecting targets, we divided our potential targets into a bright red sequence sample (Sample 1; $r<$22.5) and a faint red sequence sample (Sample 2; 22.5$<r<$23.5). We first filled our mask with as many Sample 1 targets as possible, then filled in the remainder of the mask with Sample 2 targets. While we preferentially targeted likely red sequence cluster members it was not always possible to fill the entire mask with these galaxies, in which case we would place a slit on bright blue cloud galaxies in the field. For the SDSS targeted galaxies, we selected from galaxies satisfying $z_{\text{phot}}$ within $\pm0.05(1+z_{\text{cluster}})$ of the cluster redshift and prioritized bright galaxies with a luminosity weighted selection. In these cases, Sample 2 was composed of any other bright objects outside the photometric selection. 

We used the DSIMULATOR package\footnote{\url{http://www.ucolick.org/~phillips/deimos_ref/masks.html}} to design each slit mask. DSIMULATOR automatically selects targets by maximizing the sum total weights of target candidates, by first selecting as many objects from Sample 1 as possible then filling in the remaining area of the slit mask with target candidates from Sample 2. We manually edited the automated target selection to increase the number of selected targets, e.g. by selecting another target between targets selected automatically by DSIMULATOR if the loss of sky coverage was acceptably small.

\subsubsection{Keck/DEIMOS: Data Reduction}\label{sec:KeckDeimosReduction}

The exposures for each mask were combined using the DEEP2 versions of the \emph{spec2d} and \emph{spec1d} packages \citep{Newman:2012a}. This package combines the individual exposures of the slit mosaic and performs wavelength calibration, cosmic ray removal and sky subtraction on slit-by-slit basis, generating a processed two-dimensional spectrum for each slit. The \emph{spec2d} pipeline also generates a processed one-dimensional spectrum for each slit. This extraction creates a one-dimensional spectrum of the target, containing the summed flux at each wavelength in an optimized window. The \emph{spec1d} pipeline then fits template spectral energy distributions (SED's) to each one-dimensional spectrum and estimates a corresponding redshift. There are SED templates for various types of stars, galaxies, and active galactic nuclei. We then visually inspect the fits using the \emph{zspec} software package \citep{DEEP2:2013}, assign quality rankings to each fit \citep[following a convention closely related to ][]{DEEP2:2013}, and manually fit for redshifts where the automated pipeline failed to identify the correct fit. The highest quality galaxy spectra (Q=4) have a mean signal-to-noise-ratio (SNR) of 10.7 per pixel, while the minimum quality galaxy spectra used on our redshift analysis (Q=3) have a mean SNR of 4.9 per pixel. Note that the SNR estimates are dominated by the continuum of a spectroscopic trace and an emission line galaxy may be of high quality but very low mean SNR (for example the mean SNR of a Q=4 emission line galaxy is 1.2 despite detection of H$\alpha$ and H$\beta$ or [\ion{O}{3}] in most cases). An example of one of the reduced spectra is reprinted from Figure 2 of \citet{Dawson:2015} in Figure \ref{fig:SpecCoverage} and more are shown in a related galaxy evolution paper \citet{Sobral:2015}.

In Table \ref{tab:spectra}, we present $\sim$5800 high quality galaxies and stars from our spectroscopic survey along with matched photometry from our photometric survey. 

\subsubsection{Archival Spectroscopy}

To augment our spectroscopic survey, we completed a detailed literature review of published spectroscopic redshifts of cluster members for the 29 systems in the ensemble. We compiled spectroscopic galaxies in each field using using the NASA/IPAC Extragalactic Database\footnote{\url{http://ned.ipac.caltech.edu/}} (NED). For each system we considered galaxies within 5 Mpc of the cluster center and within $\pm$10,000 km s$^{-1}$ of the mean cluster redshift to be sufficiently plausible members. Many galaxies published in the literature also appear in NED, so we cross matched and eliminated duplicate galaxies and prioritized originally published galaxies over NED matches. 

We combine all known redshifts (from NED, the literature, and our DEIMOS survey) in the cluster fields and check for duplicates using the Topcat \citep{topcat} software using the \emph{sky} function with a 1$\arcsec$\ tolerance. These combined catalogs of unique spectroscopically confirmed objects are studied in \S\ref{sec:analysis}. In Table \ref{tab:archival_spectra}, the numbers of spectroscopic redshifts from the literature review and DEIMOS survey are reported.

\begin{table*}
\begin{center}
\small
\caption{Breakdown of spectroscopy from our DEIMOS and the literature}
\begin{tabular}{lccl}
Cluster		& Unique DEIMOS	& Unique Literature 	& References\\
\hline
1RXSJ0603	& 387		& 0			& ---\\
A115 		& 255		& 76			& B83, Z90, B07, 2MASS, SDSS\\
A521 		& 0		& 193			& M00, F03\\
A523 		& 268		& 61			& G16\\
A746 		& 94		& 6			& 2MASS, SDSS\\
A781 		& 0		& 875			& G05, SDSS\\
A1240 		& 197		& 151			& B09, 2MASS, SDSS\\
A1300		& 0		& 270			& P97, Z12\\
A1612 		& 83		& 39			& SDSS\\
A2034		& 130		& 129			& SDSS, O14\\
A2061		& 0		& 404			& SDSS\\
A2163		& 0		& 407			& M08\\
A2255		& 0		& 406			& SDSS\\
A2345		& 0		& 103			& B10\\
A2443		& 253		& 17			& SDSS\\
A2744		& 0		& 695			& C87, B06, O11\\
A3365		& 313		& 33			& K98, 6dF\\
A3411		& 550		& 0			& vW17\\
CIZAJ2242	& 447		& 0			& D15\\
MACSJ1149	& 0		& 591			& SDSS, E14\\
MACSJ1752	& 432		& 0			& ---\\
PLCKG287	& 666*		& 0			& ---\\
PSZ1G108	& 290		& 0 			& ---\\
RXCJ1053	& 232		& 144			& SDSS\\
RXCJ1314	& 286		& 18			& V02\\
ZwCl0008	& 278		& 0			& G17\\
ZwCl1447	& 212		& 0			& ---\\
ZwCl1856	& 214		& 0			& ---\\
ZwCl2341	& 324		& 62			& SDSS, B13\\
\end{tabular}
\tablecomments{Reference codes in Column 4: B83$=$\citet{Beers:1983}, Z90$=$\citet{Zabludoff:1990}, B07$=$\citet{Barrena:2007}, 2MASS$=$\citet{2mass}, SDSS$=$\citet{sdss12}, M00$=$\citet{Maurogordato:2000}, F03$=$\citet{Ferrari:2003}, G16$=$\citet{Girardi:2016}, G05$=$\citet{Geller:2005}, B09$=$\citet{Barrena:2009}, P97$=$\citet{Pierre:1997}, Z12$=$\citet{Ziparo:2012}, O14$=$\citet{Owers:2014}, M08$=$\citet{Maurogordato:2008}, B10$=$\citet{Boschin:2010}, C87$=$\citet{Couch:1987}, B06$=$\citet{Boschin:2006}, O11$=$\citet{Owers:2011}, K98$=$\citet{Katgert:1998}, 6dF$=$\citet{6dF}, vW17$=$\citet{vanWeeren:2017}, D15$=$\citet{Dawson:2015}, E14$=$\citet{Ebeling:2014}, V02$=$\citet{Valtchanov:2002}, G17$=$\citet{Golovich:2017},  B13$=$\citet{Boschin:2013}}
\tablecomments{* 317 unique redshifts were obtained from VLT VIMOS Obs ID. 094.A-0529, PI M. Nonino}
\label{tab:archival_spectra}
\end{center}
\end{table*}


\section{Redshift Analysis}\label{sec:analysis}

In this section we describe the process of selecting spectroscopic cluster members from our combined redshift catalogs (see Table \ref{tab:spec} and \ref{tab:archival_spectra}). 

\subsection{Spectroscopic Catalog Generation}\label{subsec:spec_catalog}

We cut each spectroscopic catalog to only include objects within $R_{200}$ in projected space and to within $\bar{v}\pm3\sigma_{v}$, where $\bar{v}$ is the average line of sight velocity and $\sigma_{v}$ is the cluster velocity dispersion. This is accomplished with an iterative process starting with 5 Mpc and 10,000 km s$^{-1}$ and shrinking the radius and velocity window until an equilibrium catalog is achieved. This reduces the chance of inclusion of galaxies that are uninvolved in the merger. An instructive example is Abell 2061, where Abell 2067 is $\sim2.7$ Mpc (30$\arcmin$) to the northeast and at a similar redshift, but uninvolved in the merger. The iterative shrinking aperture was able to eliminate galaxies from A2067 from the redshift catalog despite being at a similar redshift because it is outside of $R_{200}$. A second example is Abell 523 ($z\sim0.1$), which has two background groups at $z\sim0.14$ within $R_{200}$ in projection \citep{Girardi:2016}. 

\subsection{One Dimensional Redshift Analysis}\label{subsec:one_dim}

\begin{figure}[ht!]
\includegraphics[width=\columnwidth]{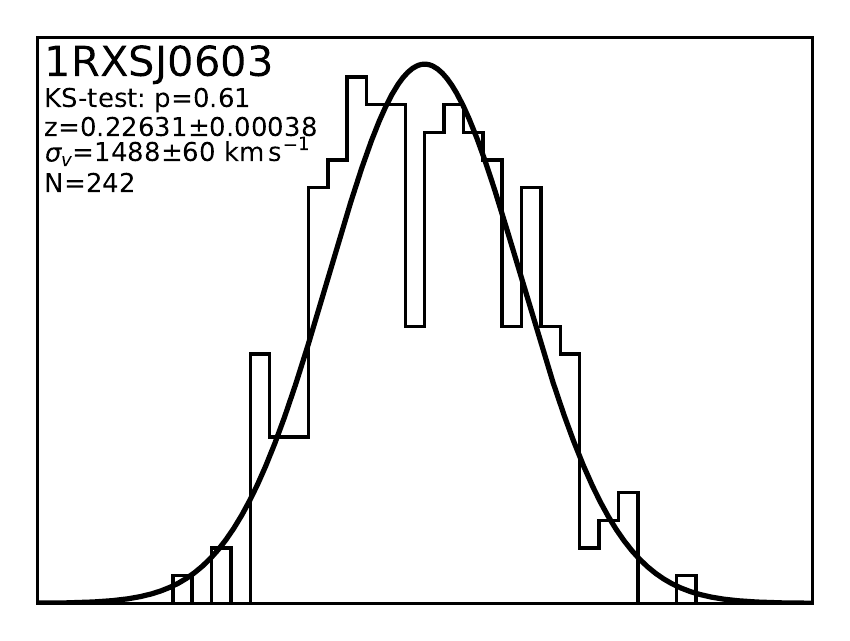}
\caption{Redshift distribution for 1RXSJ0603 based on our DEIMOS spectroscopic survey. Galaxies are selected with a shrinking 3D aperture until a stable set of galaxies within R$_{200}$ and $\pm3\sigma_{v}$ is achieved. The global redshift analysis using the biweight statistic and bias corrected 68\% confidence limits are presented in the panel. The p-value for a KS-test for Gaussianity is presented as well. The panel width is 12,000 km s$^{-1}$ centered on the cluster redshift. Bins are 300 km s$^{-1}$ at the cluster redshift. The analogous distributions for the other 28 systems in our sample are located in the appendix.}
\label{fig:hist}
\end{figure}

We display the one dimensional redshift distribution for 1RXSJ0603 in Figure \ref{fig:hist}. An analogous figure for the remaining 28 systems is presented the appendix. The corresponding normalized Gaussian distribution is overlaid with the cluster redshift and velocity dispersion given by the biweight and bias corrected 68$\%$ confidence intervals. We implement the biweight statistic based on 10,000 bootstrap samples of the member galaxies and calculate the bias-corrected 68\% confidence limits for the redshift and velocity dispersion from the bootstrap sample. This method is more robust to outliers than the dispersion of the Gaussians generated by our statistical model \citep{Beers1990}. 

We test the goodness of fit of the corresponding Gaussian distribution using a Kolmogorov-Smirnov (KS) test. The results of this analysis is displayed in Figure \ref{fig:hist} and in the Appendix for the other systems. We generally find good agreement between the spectroscopic data and single Gaussian distributions, which implies that the merging subclusters have line of sight velocity differences that are small compared to the velocity dispersion. The lowest p-value for the KS test is 0.007 for Abell 781, which is known to be composed of several subclusters with large velocity differences \citep{Geller:2005}. 

We also fit increasing numbers of Gaussians to the one-dimensional redshift distributions of each cluster utilizing an expectation-maximization Gaussian mixture model (EM-GMM) method from the Sci-Kit Learn python module. We varied the number of Gaussians from one to seven for each cluster. A one Gaussian model was strongly preferred for 27 of the 29 clusters according to the Bayesian Information Criterion (BIC). For A3365, the one Gaussian model was only slightly favored over a two Gaussian model, and for A781, a two halo model was preferred strongly. 

In a second paper (Golovich et al., in preperation), we will study the three-dimensional distribution of galaxies. 

\section{Discussion}\label{sec:discussion}

\subsection{Analysis of the Spectroscopic Survey}

In \S\ref{sssec:KeckTargetSelection} we discuss our methods of selecting targets for our spectroscopic survey. Because our photometric survey was ongoing during this process, we utilized the best available photometry for spectroscopic targeting (see Table \ref{tab:spec}). Here we analyze the success of the various targeting methods. Broadly, two distinct methods for selecting potential targets were implemented. For 21 of 54 slit masks, potential targets were identified via a photometric redshift selection based on SDSS photometric redshifts. For the remaining 33 of 54 slit masks, a red sequence selection was implemented; however, the quality of the imaging (seeing and depth) varied substantially depending on the source. In Table \ref{tab:spec}, the spectra are broken down by individual slit mask, targeting method, imaging used for targeting, and redshift. 

The biggest indication of the effect of the target imaging quality on the spectroscopic survey is with the fraction of targeted objects that yielded a secure redshift of a cluster member, which was our primary goal. In Table \ref{tab:spectra_quality}, the $\sim$7000 targeted objects are broken down by the type of object detected. Across the survey, 77\% of all targeted objects yielded a secure redshift estimate. Of these, 49\% were cluster galaxies. The largest sources of contaminants were background galaxies (26\%) and stars (18\%). 

Background galaxies detected at higher frequency with photometric redshift targeting. Detection of these objects also decreased with the redshift of the cluster. A substantial fraction of stars were detected in a few fields that had either sub-par imaging for target selection, or have low galactic latitude. The effect of imaging quality with regards to stellar contamination is evident in the five A3411 slit masks. The first four were observed using WFC targeting, and the fraction of stars increased for successive slit masks as the best member candidates were depleted by earlier masks. For the fifth slit mask, Subaru/SuprimeCam was utilized for targeting, and the fraction of stars decreased substantially. The trade off was a larger fraction of background galaxies, which is explained by the increased depth of the imaging. One benefit of the background galaxy redshift determination is for developing training sets for weak lensing source selection. Furthermore, gravitational lensing is complicated by massive structures along the line of sight, and over densities of background galaxies may help discover these types of massive background structures; however, the spectroscopic survey was not designed to detect such systems, so any detection will be serendipitous. Foreground galaxies accounted for only 6\% of secure redshifts, and these were predominately detected in higher redshift cluster fields. Finally, 305 objects were detected serendipitously; i.e., a single slit had one or more traces in addition to the targeted object. These were predominately detected in low galactic latitude fields and were composed of stars; although, $\sim$50 cluster galaxies were detected in this manner across the survey. 

\begin{table*}[h!]
\begin{center}
\scriptsize
\caption{Breakdown of detected objects for DEIMOS spectroscopic survey}
\begin{tabular}{lccccccc}
Slitmask 		& \% Secure	& \% Stars	& \% Cluster	& \% Foreground	& \% Background 	& \# Serendips\\
\hline
1RXSJ0603-1	& 88			& 14			& 59			& 3				& 11				& 14 	\\
1RXSJ0603-2	& 86			& 13			& 59			& 1				& 13				& 7   	\\
1RXSJ0603-3	& 88			& 11 			& 64			& 2				& 10				& 15 	\\
1RXSJ0603-4	& 87 			& 22			& 51			& 5				& 10				& 11 	\\
A115-1		& 78 			& 7			& 48			& 9				& 15				& 2 	\\
A115-2		& 80			& 4			& 42			& 6 				& 30				& 3 	\\
A523-1		& 82			& 2 			& 49			& 3				& 27				& 10 	\\
A523-2		& 80			& 1			& 37 			& 3				& 38				& 6 	\\
A523-3		& 83			& 5			& 34			& 1				& 43				& 5	\\
A746-1		& 87			& 2			& 60			& 5				& 22				& 2	\\
A1240-1		& 72			& 2 			& 41			& 3				& 25				& 5	\\
A1240-2		& 68 			& 4			& 27			& 4				& 33				& 3	\\
A1612-1		& 50			& 6			& 30			& 3				& 10				& 5	\\
A2034-1		& 80 			& 2			& 39			& 3				& 37				& 3	\\
A2443-1		& 83			& 1			& 58			& 1				& 24				& 5	\\
A2443-2		& 79			& 3			& 39 			& 5				& 31				& 7	\\
A3365-1		& 91			& 9			& 49			& 0				& 31				& 2	\\
A3365-2		& 87 			& 12			& 39			& 0				& 35				& 4	\\
A3365-3		& 79 			& 6			& 33			& 0				& 40				& 2	\\
A3365-4		& 61			& 3			& 14			& 0				& 45				& 1	\\
A3411-1		& 77 			& 25			& 46			& 0				& 5		 		& 1	\\
A3411-2		& 81			& 43			& 25			& 1				& 12				& 4 	\\
A3411-3		& 93			& 28			& 43			& 0				& 11				& 2	\\
A3411-4		& 90 			& 58			& 20			& 0				& 12		 		& 2	\\
A3411-5		& 76 			& 12			& 37			& 0				& 27		 		& 2	\\
CIZAJ2242-1	& 74			& 18			& 49			& 2				& 4		 		& 12	\\
CIZAJ2242-2	& 77 			& 24			& 48			& 2				& 2		 		& 25	\\
CIZAJ2242-3	& 79			& 36			& 29			& 7				& 8		 		& 23	\\
CIZAJ2242-4	& 86			& 25			& 50			& 3				& 7		 		& 22	\\
MACSJ1752-1	& 81			& 8			& 47			& 14				& 15				& 3	\\
MACSJ1752-2	& 84 			& 8			& 49			& 14				& 13				& 4	\\
MACSJ1752-3	& 83			& 9			& 32 			& 13				& 29				& 9	\\
MACSJ1752-4	& 84			& 8			& 41			& 17				& 18				& 2	\\
PLCKG287-1	& 69 			& 0			& 47			& 23				& 7		 		& 1	\\
PLCKG287-2	& 71 			& 3			& 45			& 12				& 10				& 2	\\
PLCKG287-3	& 41			& 3			& 18			& 12				& 8		 		& 0	\\
PSZ1G108-1	& 75 			& 53			& 14			& 6				& 2				& 2	\\
PSZ1G108-2	& 85			& 73			& 8			& 2				& 2				& 2	\\
RXCJ1053-1	& 72			& 0			& 17 			& 2				& 53				& 2 	\\
RXCJ1053-2	& 89			& 7			& 38			& 0				& 44				& 0	\\
RXCJ1053-3	& 77			& 1			& 27			& 1				& 46				& 1 	\\
RXCJ1314-1	& 79 			& 4			& 45			& 6				& 24				& 3	\\
RXCJ1314-2	& 64 			& 0			& 27			& 4				& 32				& 3	\\
ZwCl0008-1	& 74			& 6			& 53			& 0				& 14				& 5	\\
ZwCl0008-2	& 79 			& 23			& 38	 		& 0				& 16				& 10 \\
ZwCl0008-3	& 77 			& 23			& 32			& 0				& 23				& 14	\\
ZwCl0008-4	& 89 			& 27			& 19			& 4				& 37				& 17	\\
ZwCl1447-1	& 77 			& 1 			& 49			& 13				& 14				& 1	\\
ZwCl1447-2	& 65			& 2			& 30			& 16				& 16				& 3	\\
ZwCl1856-1	& 83			& 54			& 22			& 5				& 4				& 1	\\
ZwCl1856-2	& 85 			& 64			& 15			& 5				& 1				& 3	\\
ZwCl2341-1	& 71			& 0			& 43			& 4				& 23				& 7	\\
ZwCl2341-2	& 77 			& 2			& 46			& 6				& 23				& 4	\\
ZwCl2341-3	& 82 			& 2			& 44			& 5				& 30		 		& 1	\\
\hline \hline
Targeting Method & & & & & &\\
 \hline
Photometric Redshifts	&  76			& 4			& 	41		& 		7		& 		25		& 72	\\
Color--Magnitude		&  77			& 21			& 	35		& 		4		& 		16 		& 233	\\
\hline \hline
Imaging	&&&&&&\\
\hline
WFC			& 83			& 20			& 43			& 1				& 18				& 168\\
SDSS		& 76			& 4			& 41			& 7				& 25				& 72    \\
SC			& 67			& 7			& 35			& 8				& 19				& 57	\\
DSS			& 81			& 60			& 14			& 5				& 2				& 8    \\
\hline \hline
Cluster Redshifts & & & & & & \\
\hline
$z<0.1$		& 	76		& 	5		& 		28	& 		0		& 	43			&   12  \\
$0.1<z<0.2$	& 	78		& 	15		& 		39	& 		3		& 	20			& 193	\\
$0.2<z<0.3$	& 	79		& 	6		& 		47	& 		4		& 	22			&   67	\\
$z>0.3$		& 	74		& 	22		& 		32	& 		12		& 	10			&   33  \\
\hline \hline
Totals		& 77			& 14			& 38			& 5				& 20				& 305 
\end{tabular}
\label{tab:spectra_quality}
\tablecomments{Percentages of stars, cluster members, foreground and background objects may not add to the total percentage of secure objects due to rounding. Column 1: cluster and slit mask number; Column 2: percentage of secure redshifts among targeted objects; Column 3-6: percentage of stars, cluster member galaxies, foreground galaxies, and background galaxies, respectively; Column 7: number of serendipitous detections.} 
\end{center}
\end{table*}

\subsection{Cluster Redshift Histograms}

The presence of radio relics in merging galaxy clusters constitutes a strong prior for ongoing merging activity. Given this, these 29 merging clusters are expected to be composed of two or more subclusters. However, 28 of the 29 systems are well fit by a single Gaussian ($p>0.05$). There are two potential explanations for this, which are not mutually exclusive: 1) radio relics indicate a merger occurring within the plane of the sky (transverse to the line of sight), and/or 2) radio relics indicate a merger observed near apocenter. 

Based on the redshift results along, both scenarios are plausible. First, most of our relics were detected in shallow surveys, and the surface brightness is higher when the line of sight intersects a large fraction of the emission in three dimensions \citep{Skillman:2013}. Furthermore, detected radio relics have been shown to be highly polarized \citep[e.g][]{Govoni:2004b,Ferrari:2008}, which correlates with a transverse viewing angle \citep{ensslin1998}. Second, radio relics occur for only a small fraction of the full merger phase, and it takes time for the radio relic to develop \citep[see Figure 5 of][]{Skillman:2013}. This may explain why the Bullet Cluster's bow shock is not coincident with a bright radio relic. Meanwhile, El Gordo contains radio relics, and it was shown to be returning from apocenter \citep{Ng:2015}. Thus, it is likely a combination of the two scenarios that explain the unimodal redshift distributions in 28 of 29 systems in the sample. However, recent magneto-hydrodynamical simulations suggest that explanation (1) is more likely for radio relic systems \citep{Vazza:2012,Wittor:2017}

The one outlier, A781, is known to be composed of multiple clusters at various redshifts \citep{Geller:2005}. The system is composed of two clusters in projection at $z\sim0.3$ and $z\sim0.4$. Here we studied the $z\sim0.3$ system, which is further split into two redshift peaks (see Figure \ref{fig:A2}). The radio relic is associated with the slightly higher redshift peak on the western side of the cluster. The lower redshift peak is associated with an infalling subcluster, which is yet to merge, based on the undisturbed X-ray surface brightness distribution \citep[see Figure 1 of][where this subcluster is referred to as the Middle subcluster]{Sehgal08}. 

\subsection{Potential Uses for These Data}
 
 The photometric data presented in this paper are sufficiently deep for detailed, wide field, weak gravitational lensing analyses of each cluster \citep[see][for a review]{Bartelmann:2001, Hoekstra:2013}. This will allow for the mapping of the total mass distribution of each system, as well as allow for mass estimation of each subcluster in a manner unbiased by the dynamical interaction of the merger \citep[e.g.,][]{Jee:2015,Jee:2016}. Both galaxy velocity dispersion based mass estimates assuming the system is virialized \citep{Takizawa:2010} and X-ray temperature or luminosity scaling relation based mass estimates assuming the cluster is in hydrostatic equilibrium \citep{Zhang:2010} overestimate the mass in merging systems; although, \citet{Takizawa:2010} show that mergers along the line of sight are more strongly affected by this, and most strongly near core passage. Radio relic systems are typically observed $\sim$1 Gyr after pericentric passage \citep{Ng:2015,Golovich:2016,Golovich:2017,vanWeeren:2017}.
 
\begin{figure}
\begin{center}
\includegraphics[width=\columnwidth]{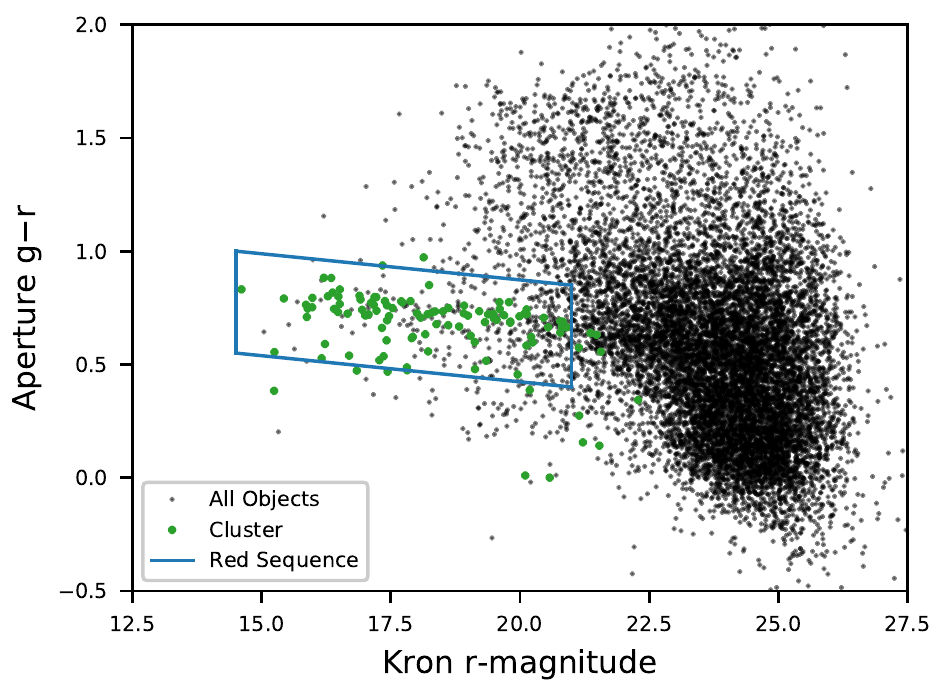}
\caption{Color--magnitude diagram for the photometric catalog of RXCJ1053 with overlaid spectroscopic matches. The red sequence selection box is shown in blue.}
\label{fig:sample_redsequence}
\end{center}
\end{figure}

Since we took images with two photometric filters that straddle the 4000\AA\, break, colors may be assigned to objects allowing color magnitude selection. This allows for selection of background galaxies for weak lensing as well as cluster members based on the red sequence technique (see Figure \ref{fig:sample_redsequence}). 
 
The spectroscopic data contains the added information of line of sight motion, which allows for a pure catalog of cluster members. From this catalog, dynamical modeling of the mergers may be achieved. Merging clusters are efficient astrophysical laboratories for studying several phenomenon including particle acceleration, cool-core disruption, and potential self-interacting dark matter signals. Many of these are time and velocity dependent, which require accurate dynamical models to fully understand. Furthermore, these dynamical models are invaluable for simulators in the form of constrained initial conditions. Finally, the spectral quality from DEIMOS allows for analyses of merging induced star formation, galaxy evolution, and AGN activity \citep[see e.g.,][]{Sobral:2015}.

\subsection{Summary}
In this paper, we presented our observational strategy, reduction, and analysis of $\sim$20 hours of Subaru/SuprimeCam imaging of 29 merging galaxy clusters alongside our spectroscopic follow up of 7000 objects (54 slit masks) with Keck/DEIMOS. We presented $\sim$5800 new high quality galaxy and star spectra from our spectroscopic survey matched to the photometry from our Subaru/SurpimeCam survey in Table \ref{tab:spectra}. These data are combined with literature spectroscopy and SuprimeCam imaging, which resulted in $\sim$5400 cluster members in total across the 29 systems. A one dimensional redshift analysis showed that 28 of 29 of the systems are well fit by a single Gaussian distribution. This suggests the ongoing mergers are occurring either within the plane of the sky or are observed near apocenter (or a combination of the two factors). We analyzed the effect of different imaging sources and selection methods for targeting slits in our spectroscopic survey, and we discussed possible uses for this large data set of photometric and spectroscopic observations of galaxies within merging galaxy clusters.
\\
\section{Acknowledgments}
We would like to thank the broader membership of the Merging Cluster Collaboration for their continual development of the science motivating this work, for useful conversations, and for diligent proofreading, editing, and feedback. 
This material is based upon work supported by the National Science Foundation under Grant No. (1518246).
This material is based in part upon work supported by STSci grant HST-GO-13343.001-A.
Part of this was work performed under the auspices of the U.S. DOE by LLNL under Contract DE-AC52-07NA27344.
Some of the data presented herein were obtained at the W.M. Keck Observatory, which is operated as a scientific partnership among the California Institute of Technology, the University of California and the National Aeronautics and Space Administration. The Observatory was made possible by the generous financial support of the W.M. Keck Foundation.
Based in part on data collected at Subaru Telescope, which is operated by the National Astronomical Observatory of Japan.
Funding for the Sloan Digital Sky Survey IV has been provided by the Alfred P. Sloan Foundation, the U.S. Department of Energy Office of Science, and the Participating Institutions. SDSS acknowledges support and resources from the Center for High-Performance Computing at the University of Utah. The SDSS web site is www.sdss.org.
The Digitized Sky Surveys were produced at the Space Telescope Science Institute under U.S. Government grant NAG W-2166.
Funding for the DEEP2/DEIMOS pipelines has been provided by NSF grant AST-0071048. 
The DEIMOS spectrograph was funded by grants from CARA (Keck Observatory) and UCO/Lick Observatory, a NSF Facilities and Infrastructure grant (ARI92-14621), the Center for Particle Astrophysics, and by gifts from Sun Microsystems and the Quantum Corporation.
This research has made use of the NASA/IPAC Extragalactic Database (NED) which is operated by the Jet Propulsion Laboratory, California Institute of Technology, under contract with the National Aeronautics and Space Administration.
This research has made use of NASA's Astrophysics Data System.
Based in part on data collected at Subaru Telescope and obtained from the SMOKA, which is operated by the Astronomy Data Center, National Astronomical Observatory of Japan.

{\it Facilities:} \facility{Keck (DEIMOS)} \facility{INT (WFC)} \facility{Subaru (SuprimeCam)} \facility{VLT (VIMOS)}

\appendix
\setcounter{table}{0}
\setcounter{figure}{0}
\renewcommand\thefigure{\thesection.\arabic{figure}}    
\renewcommand\thetable{\thesection.\arabic{table}}    

\section{A. Spectroscopic Catalog}
Table \ref{tab:spectra} contains the R.A. and DEC. coordinates, redshifts, Subaru/SuprimeCam magnitudes, and spectral features for 5594 galaxies and stars identified by our DEIMOS spectroscopic survey (see \S\ref{subsec:DEIMOS_observations}). Each spectroscopically confirmed object was matched with the Subaru/SuprimeCam catalog (see \S\ref{ssec:photometry}) using the Topcat software \citep{topcat} with a 1$\arcsec$ tolerance. Objects without photometric matches were discarded. Photometric objects were matched to their nearest spectroscopic match and were not allowed to match more than once. 

\begin{table*}[h!]
\caption{DEIMOS Spectroscopic Survey Catalog}
\small
\begin{tabular}{lllllllll}
ID & RA & DEC & g & r & i & $\text{z}$ & $\sigma_\text{z}$ & Spectral Features \\
\hline
1 & 90.84466369 & 42.27306837 & 20.28 & 18.81 & 17.78 & 0.220011 & 3.81E-05  & Hb ab, Mg I (b), [Fe I], Na I (D), Ha ab \\
1 & 90.81274054 & 42.25876563 & 23.58 & 22.16 & 21.12 & 0.508420 & 3.06E-05  & Mg I (b), [Fe I]m Na I (D), Ha \\
1 & 90.90432650 & 42.12064156 & 21.90 & 20.36 & 19.32 & 0.224067 & 3.92E-05  & G band, Hb ab, Mg I (b), [Fe I] \\
1 & 90.84777475 & 42.17517749 & 21.20 & 19.71 & 18.71 & 0.225441 & 3.92E-05  & G band, Mg I (b), [Fe I], Na I (D) \\
1 & 90.80537365 & 42.16394040 & 22.49 & 21.05 & 20.08 & 0.227767 & 3.99E-05  & Hb ab, Mg I (b), Na I (D), Ha ab \\ 
\end{tabular}
\label{tab:spectra}
\tablecomments{Table A.1 is published in its entirety in the machine-readable format. A portion is shown here for guidance regarding its form and content. Column 1: Cluster ID (1$=$1RXSJ0603, 2$=$A115, 3$=$A523, 4$=$A746, 5$=$A1240, 6$=$A1612, 7$=$A2034, 8$=$A2443, 9$=$A3365, 10$=$A3411, 11$=$CIZAJ2242, 12$=$MACSJ1752, 13$=$PLCKG287, 14$=$PSZ1G108, 15$=$RXCJ1053, 16$=$RXCJ1314, 17$=$ZwCl0008, 18$=$ZwCl1447, 19$=$ZwCl1856, 20$=$ZwCl2341); Column 2: Right Ascension (J2000); Column 3: Declination (J2000); Column 4: g band magnitude; Column 5: r band magnitude; Column 6: i band magnitude; Column 7: Redshift; Column 8: Redshift Uncertainty; Column 9: Spectral Features Identified in 1D Spectrum} 
\end{table*}

\section{B. Spectroscopic Redshift Histograms}

\begin{figure*}[h!]
\includegraphics[width=\textwidth]{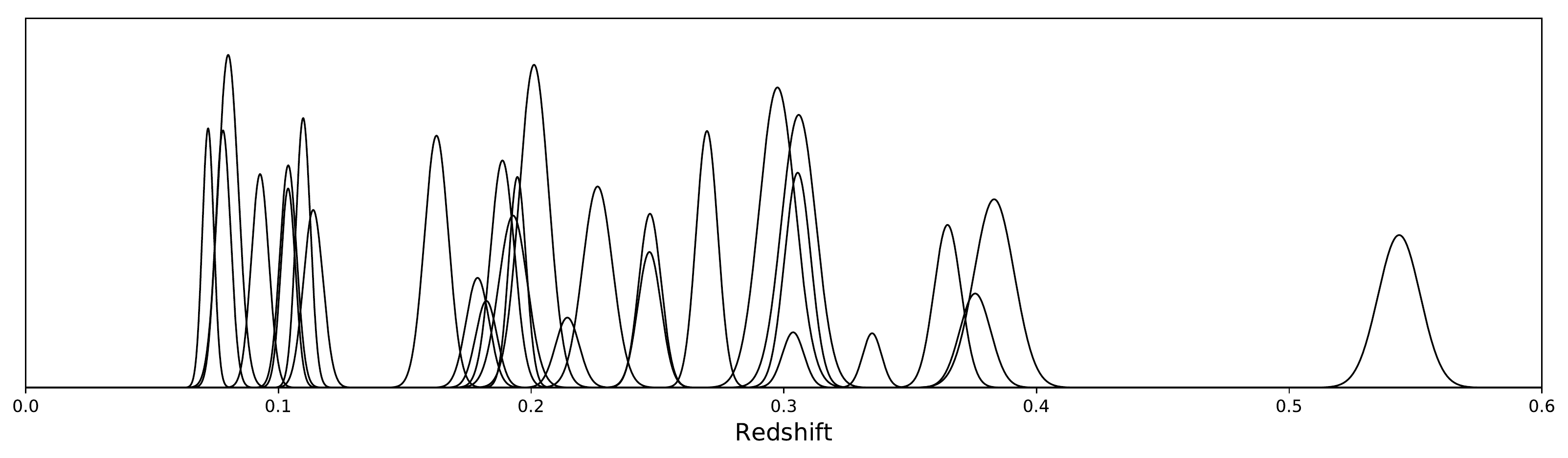}
\caption{Redshift distributions for 29 radio relic selected merging galaxy clusters. Gaussians are generated based on the biweight and bias-corrected 68$\%$ confidence interval from 10,000 bootstrap realizations of the spectroscopic catalog for each system. Gaussians are scaled to be proportional to the number of spectroscopic galaxies for each system.}
\label{fig:allspec}
\end{figure*}

In \S\ref{sec:analysis} we compiled the spectroscopic cluster member catalogs using an iterative, shrinking three dimensional aperture method. The resulting galaxies were then fit with a one dimensional Gaussian with a biweight and bias-corrected confidence interval analysis. In Figure \ref{fig:allspec}, the 29 resulting Gaussians are presented to demonstrate the sample distribution of spectroscopic cluster members. The area of a given Gaussian is proportional to the population of galaxies in the respective cluster catalogs. 

\begin{figure}
\begin{center}
\includegraphics[width=\textwidth]{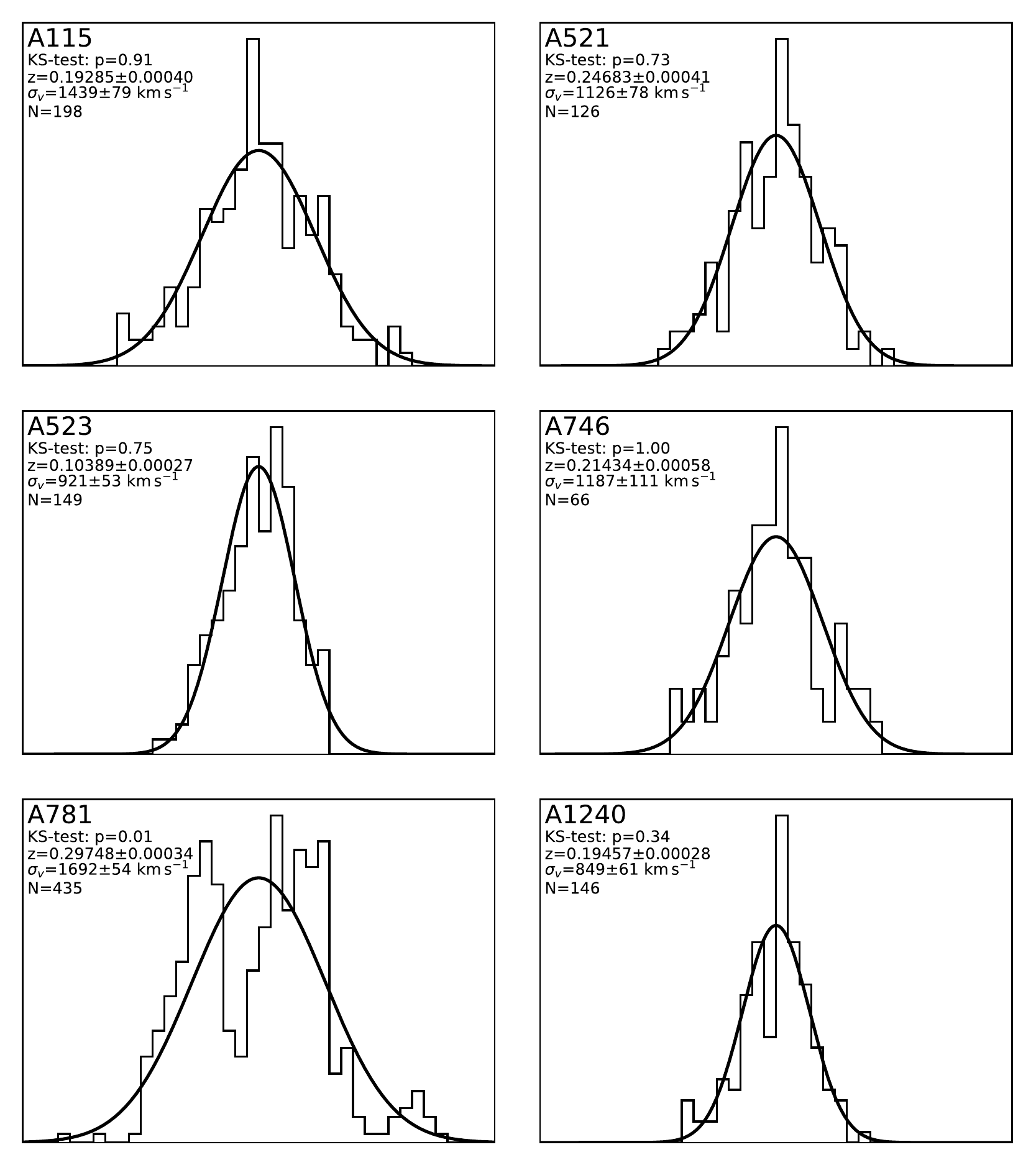}
\caption{Analogous redshift distributions to Figure \ref{fig:hist} for remaining systems in the ensemble.}
\label{fig:A2}
\end{center}
\end{figure}

In Figures \ref{fig:A2}, the analogous redshift distributions to that presented in Figure \ref{fig:hist} for 1RXSJ0603 are presented for the remaining 28 systems. \\

\addtocounter{figure}{-1}
\begin{figure}
\begin{center}
\includegraphics[width=\textwidth]{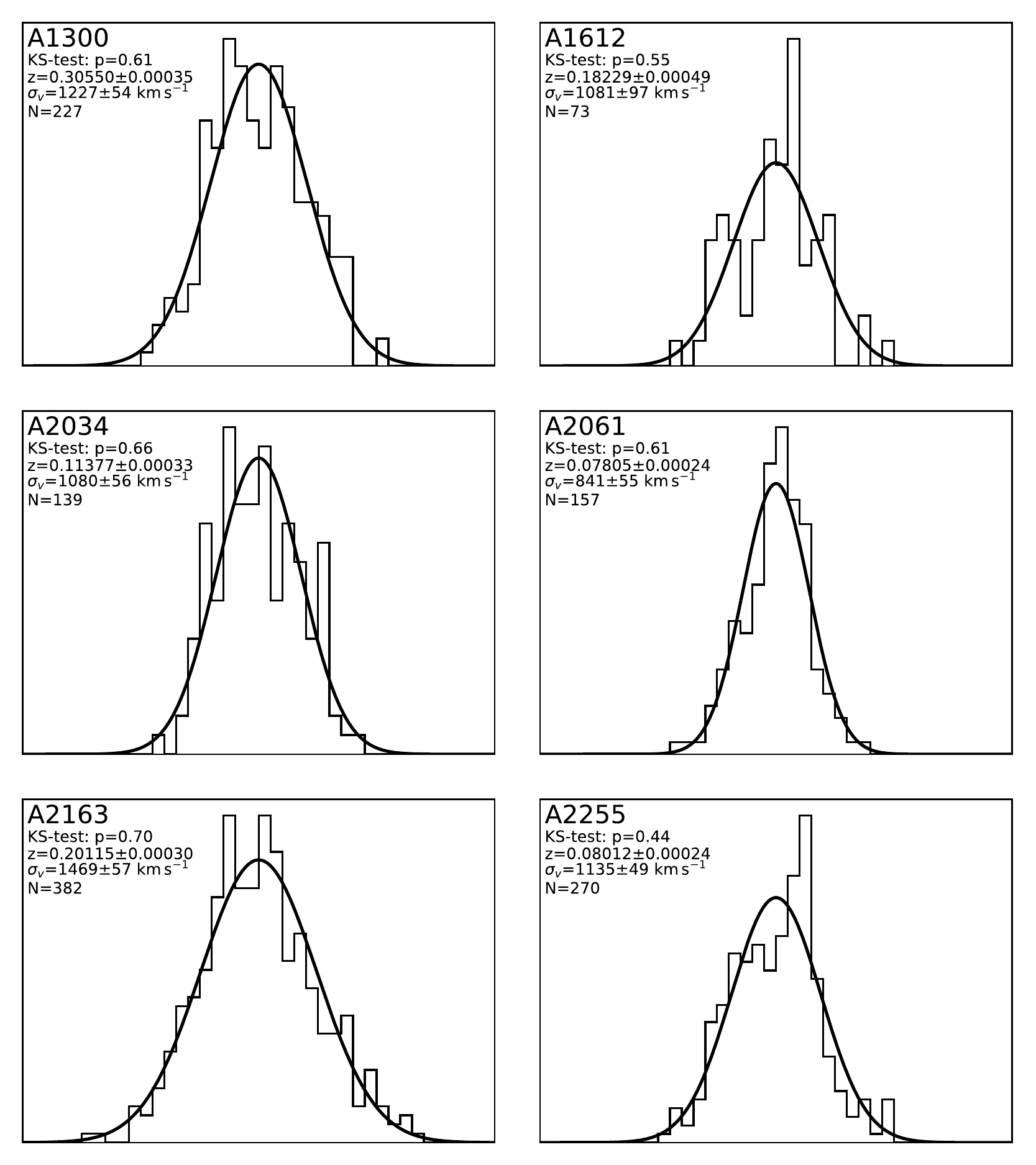}
\caption{\textbf{(Figure \ref{fig:A2} Continued)} Analogous redshift distributions to Figure \ref{fig:hist} for remaining systems in the ensemble.}
\end{center}
\end{figure}

\addtocounter{figure}{-1}
\begin{figure}
\begin{center}
\includegraphics[width=\textwidth]{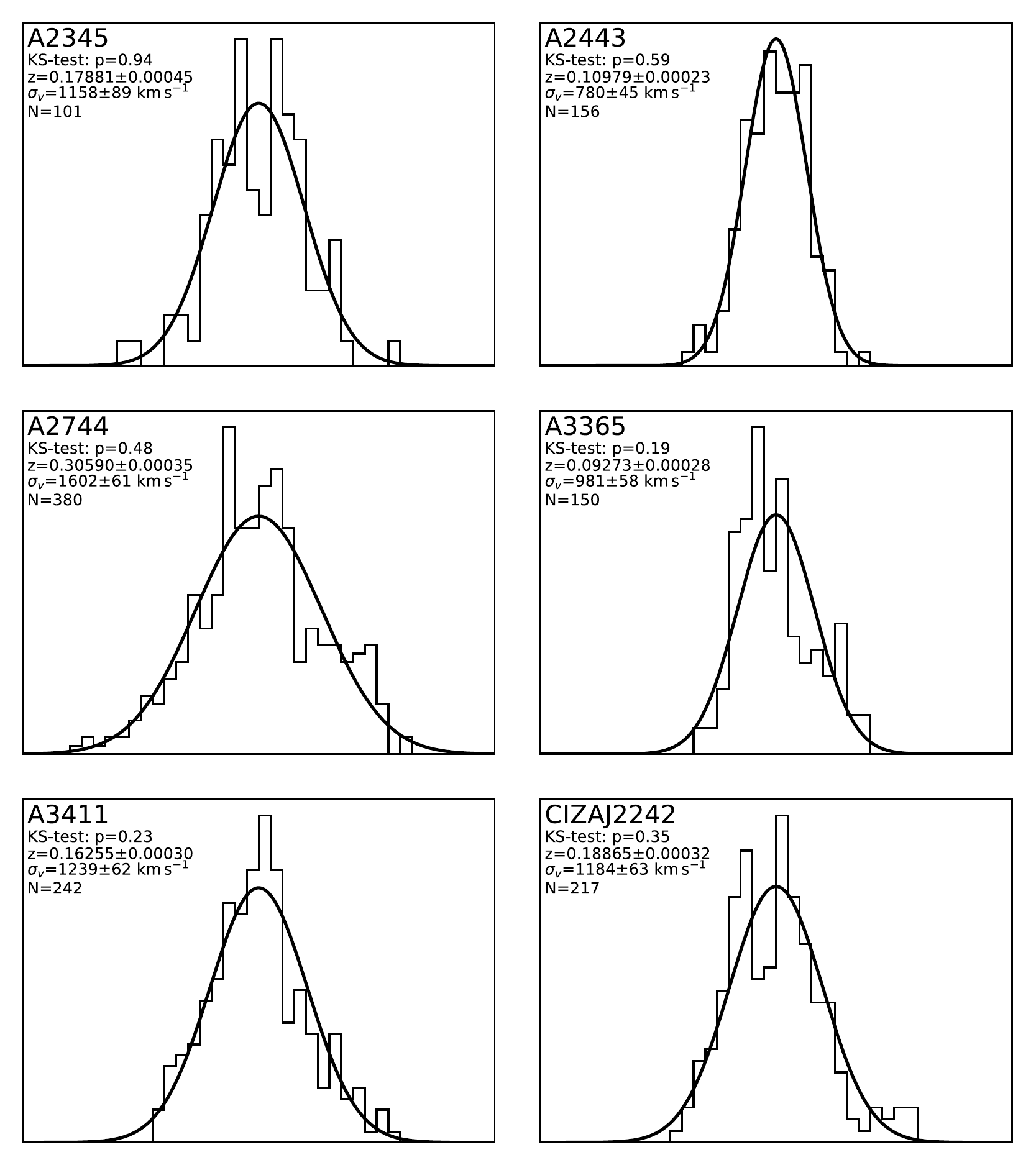}
\caption{\textbf{(Figure \ref{fig:A2} Continued)} Analogous redshift distributions to Figure \ref{fig:hist} for remaining systems in the ensemble.}
\end{center}
\end{figure}

\addtocounter{figure}{-1}
\begin{figure}
\begin{center}
\includegraphics[width=\textwidth]{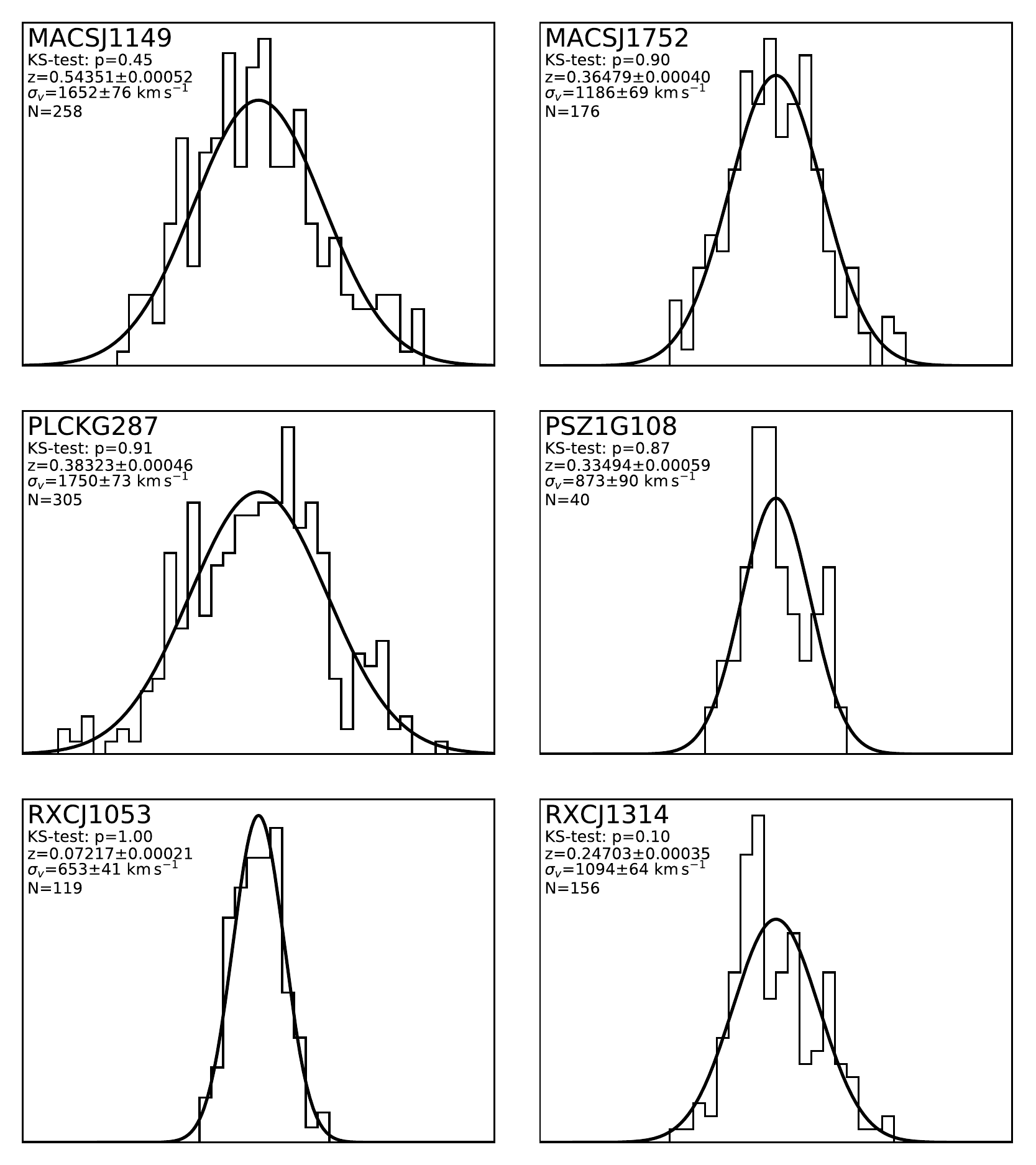}
\caption{\textbf{(Figure \ref{fig:A2} Continued)} Analogous redshift distributions to Figure \ref{fig:hist} for remaining systems in the ensemble.}
\end{center}
\end{figure}

\addtocounter{figure}{-1}
\begin{figure}
\begin{center}
\includegraphics[width=\textwidth]{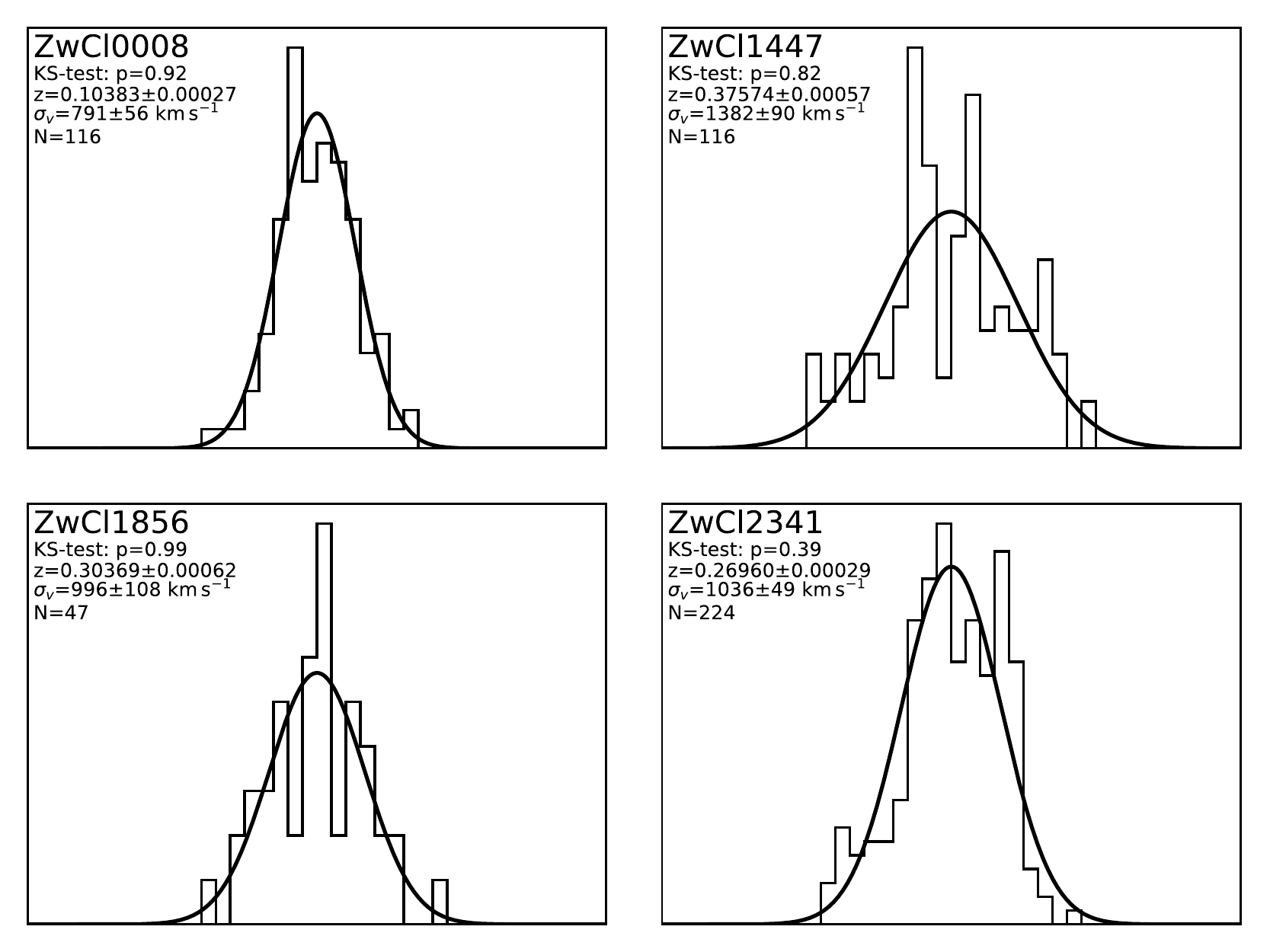}
\caption{\textbf{(Figure \ref{fig:A2} Continued)} Analogous redshift distributions to Figure \ref{fig:hist} for remaining systems in the ensemble.}
\end{center}
\end{figure}

\bibliographystyle{aasjournal.bst}
\bibliography{mcc}
\end{document}